\documentclass[11pt]{article}

\usepackage{amssymb,amsfonts,amsmath,fancybox,boxedminipage,epsfig,wrapfig}
\usepackage{psfrag,rotating}
\usepackage{graphicx}
\usepackage{cite}

\newcommand{\dket}{\rangle \! \rangle}
\newcommand{\dbra}{\langle \! \langle}
\newcommand{\ket}{\rangle}
\newcommand{\bra}{\langle}

\newcommand{\bpartial}{\bar{\partial}}
\newcommand{\hotimes}{\hat{\otimes}}
\newcommand{\non}{\nonumber}

\newcommand{\bDelta}{\bar{\Delta}}
\newcommand{\uDelta}{\underline{\Delta}}
\newcommand{\ueta}{\underline{\eta}}
\newcommand{\vpc}{\varphi_{\text{cl}}}
\newcommand{\tgamma}{\tilde{\gamma}}
\newcommand{\bmu}{\bar{\mu}}
\newcommand{\w}{\varpi} 

\newcommand{\uw}{{\underline{\varpi}}}
\newcommand{\bpsi}{{\bar{\psi}}}
\newcommand{\bomega}{{\bar{\omega}}}
\newcommand{\homega}{{\hat{\omega}}}
\newcommand{\bxi}{{\bar{\xi}}}
\newcommand{\hxi}{{\hat{\xi}}}

\newcommand{\A}{{\mathcal A}}
\newcommand{\C}{{\mathbb C}}
\newcommand{\cop}{{\rm cop}}
\newcommand{\e}{\rm \acute{e}}
\newcommand{\be}{\mathbf{e}}
\newcommand{\id}{{\rm id}}
\newcommand{\cH}{{\mathcal H}}
\newcommand{\J}{{\mathcal J}}
\newcommand{\bJ}{\bar{J}}
\newcommand{\hj}{{\hat{\jmath}}}
\newcommand{\hJ}{{\hat{J}}}
\newcommand{\hK}{{\hat{K}}}
\newcommand{\bK}{\bar{K}}
\newcommand{\K}{{\mathbb K}}
\newcommand{\cK}{{\mathcal K}}
\newcommand{\cL}{{\mathcal L}}
\newcommand{\bL}{{\bar{L}}}

\newcommand{\hM}{{\hat{M}}}

\newcommand{\n}{{\rm \tilde{n}}}
\newcommand{\N}{{\mathbb N}}

\newcommand{\CO}{{\cal O}}

\newcommand{\hp}{\hat{p}}
\newcommand{\up}{\underline{p}}
\newcommand{\hq}{{\hat{q}}}
\newcommand{\bq}{\bar{q}}

\newcommand{\CS}{{\mathcal S}}
\newcommand{\bT}{\bar{T}}

\newcommand{\Tc}{T_{\text{cl}}}
\newcommand{\uT}{{\underline{T}}}
\newcommand{\bTc}{\bar{T}_{\text{cl}}}
\newcommand{\bw}{\bar{w}}
\newcommand{\X}{\underset{{\mathcal C}(X)}}

\newcommand{\bz}{\bar{z}}
\newcommand{\Z}{{\mathbb Z}}
\newcommand{\cZ}{{\mathcal Z}}

\begin{document}

\begin{titlepage}
\vspace{.5in}
\flushright{UCD-2003-10\\
hep-th/0310234\\
October 2003}\\
\vspace*{.5cm}
\begin{center}
{\Large\bf  Quantum Liouville theory and BTZ black hole entropy}\\[2ex]

Yujun Chen \footnote{chen@physics.ucdavis.edu}\\
\vspace{.1cm}
       {\small\it Department of Physics}\\
       {\small\it University of California}\\
       {\small\it Davis, CA 95616, USA}
\end{center}
\vspace*{1cm}
\begin{center}
{\large\bf Abstract}
\end{center}
\begin{center}
\begin{minipage}{5in}
{\small 
\noindent  In this paper I give an explicit conformal field theory description of 
(2+1)-dimensional BTZ black hole
entropy.  In the boundary Liouville field theory I investigate the reducible 
Verma modules in the elliptic sector, which correspond to certain
irreducible representations of the quantum algebra $U_q(sl_2) \odot U_\hq(sl_2)$. 
I show that there are states that decouple from these reducible Verma modules in
a similar fashion to the decoupling of  null states in minimal models.  Because of
the nonstandard form of the Ward identity for the
two-point correlation functions in quantum Liouville field theory, these decoupling
states have positive-definite norms.  The explicit counting from these states gives the
desired Bekenstein-Hawking entropy in the semi-classical limit when
$q$ is a root of unity of odd order.
\vspace*{.25cm}
}
\end{minipage}
\end{center} 
\end{titlepage}

\addtocounter{footnote}{-1}

\section{Introduction}
\label{introduction}
\setcounter{equation}{0}

The classical laws of black hole mechanics together with the
temperature of Hawking radiation suggest the identification of
A/4 with the physical entropy of a black hole, where $A$ is the area
of the horizon.  A major goal of research in quantum gravity
is to understand the statistical origin of this formula.  Besides
serving as a useful model for realistic black hole physics, the
(2+1)-dimensional BTZ black hole \cite{btz} has also been found to be related
to the near-horizon geometries of many high dimensional black hole
solutions (see for example,\cite{strominger,bss,Balasubramanian,Iofa,Lee,Cvetic,Cvetic2,
Hyun,Sfetsos}).  Thus it is of strong importance to understand BTZ black
hole entropy from first principles.

The asymptotic symmetry group  \cite{bh} of (2+1)-dimensional gravity with
negative cosmological constant $\Lambda = -1/l^2$ is 
generated by two copies of the Virasoro algebra, with classical central charge
\begin{align}
  c_L = c_R = \frac{3l}{2G},
\end{align}
where $G$ is the gravitational constant in 2+1 dimension.  Based on
this fact,  a simple derivation of the BTZ black hole
entropy was given  in \cite{strominger, bss} using Cardy's formula
\cite{cardy1,cardy2,carlip1}, which states that the
asymptotic density of states for a conformal field theory is given by 
\begin{align}
\label{cardy}
  \rho (\Delta, \bar{\Delta}) 
  \sim \exp \Bigg \{ 2 \pi \sqrt{\frac{(c_R - 24 \Delta_0)\Delta}{6}}\Bigg \}  
   \exp \Bigg \{2 \pi \sqrt{\frac{(c_L - 24 \bar{\Delta}_0) \bar{\Delta} }{6}}\Bigg \},
\end{align}
where $\Delta$, $\bar{\Delta}$ are the eigenvalues of Virasoro
generators $L_0$ and $\bar{L}_0$, and $\Delta_0$,
$\bar{\Delta}_0$ the lowest eigenvalues.  For the BTZ black
hole \cite{banados, strominger},
\begin{align}
  M = (\Delta + \bDelta)/l, ~~ J= \Delta- \bDelta, 
\end{align}
where $M$ and $J$ are the mass and angular momentum of the black
hole.  Substituting into (\ref{cardy}), assuming that 
\begin{align}
\label{lowest}
  \Delta_0 =\bar{\Delta}_0 =0,  
\end{align}
we obtain the Bekenstein-Hawking entropy for BTZ black hole. 

Unfortunately, such a derivation does not tell us what microscopic
degrees of freedom contribute to the black hole entropy.  The presence
of the asymptotic conformal algebra strongly suggests that the
asymptotic dynamics is described by a two-dimensional conformal field
theory.  We thus would like to have a more concrete understanding of
the black hole entropy by explicitly counting the states in this
boundary conformal field theory.  

A good candidate for such a boundary conformal field theory is
Liouville field theory.  (2+1)-dimensional Einstein gravity with $\Lambda
<0$ can be reformulated as a Chern-Simons gauge theory with gauge
group $SL(2,R) \times SL(2,R)$ \cite{at,witten}, with gauge potentials
\begin{align}
\label{gauge}
    A^{(\pm) a} = \omega^a \pm e^a/l,
\end{align}
where $e^a = e^a_{\mu} d x^{\mu}$ is the triad and $\omega^a =
\frac{1}{2} \epsilon^{abc} \omega_{\mu bc} dx^{\mu}$ is the spin
connection.  The Einstein-Hilbert action becomes 
\begin{align}
  I = I_{\rm CS}[A^{(-)}] - I_{\rm{CS}}[A^{(+)}],
\end{align}  
where
\begin{eqnarray}
   I_{\rm CS}[A] = \frac{l}{16 \pi G} \int_{M} \text{Tr} \, \big \{ 
     A \wedge dA  + \frac{2}{3} A \wedge A \wedge A \big \}
\end{eqnarray}
is the Chern-Simons action.  The asymptotically AdS boundary condition
\cite{bh, coussaert} reduces the asymptotic dynamics
to a boundary Liouville theory \cite{coussaert, rs}.

There have been questions whether Liouville field theory provide
enough microstates for BTZ black hole entropy counting
\cite{martinec}.  This issue is related to
the spectrum of Liouville theory, which we will discuss in
Section \ref{hilbert1}.  After canonical quantization, the spectrum of Liouville
theory consists of two different classes.  In Seiberg's
notation, they are called the normalizable macroscopic states and the nonnormalizable
Hartle-Hawking states respectively.  The state-counting of the normalizable states is well
understood.  Define the effective central charge \cite{kutasov,carlip1} as
\begin{align}
  c_{\text{eff}} = c - 24 \Delta_0.
\end{align}
The lowest Virasoro eigenvalue for normalizable states is $\Delta_0
= (c-1)/24$ rather than $0$, so $c_{\text{eff}} =1$.  Thus the density of states behaves like
that of an ordinary scalar field, which
does not provide enough states for entropy counting.   However, the condition (\ref{lowest}) may
be satisfied by the
nonnormalizable Hartle-Hawking states, which
correspond to local operator insertions.  This suggests that in order to understand the state
counting, we ought to investigate these Hartle-Hawking states instead.   

The understanding of state-counting in Liouville field theory has
further motivations, since Liouville theory can also be obtained near
the horizon of an arbitrary black hole by dimensionally reducing to $r-t$ plane
\cite{solodukhin, giacomini}.  There, similarly, the classical central
charge of Liouville theory
gives the Bekenstein-Hawking entropy of the black hole by applying
Cardy's formula.  Thus the understanding of quantum Liouville
theory and its explicit state counting could offer an
explanation for the ``universality'' of the Bekenstein-Hawking entropy
in different approaches to quantum gravity.

In this paper we first review in Section \ref{sec:classical} classical Liouville
field theory, which is closely related to the description of
two-dimensional surfaces.  We follow in Section \ref{quantization} 
and \ref{sec:group} with a summary of the canonical quantization
procedure proposed by Gervais and his collaborators 
\cite{gn1,gn2,gn3,gn4,gn5,gn6,gn7,ger1,ger2,ger3},
which shows manifestly the underlying quantum algebra structure of the
theory.  In Section \ref{construction}, we construct the Hartle-Hawking
states corresponding to certain irreducible representations of the
quantum algebra.  The conformal weights of these states are of Kac
form, and  the Verma modules built on them are reducible.  
In order to define the norm of the states decoupling from these
reducible Verma modules, Section \ref{ward identity} gives a discussion of the Ward
identity of the two-point functions in Liouville field theory, whose
difference with the standard form has a geometric
origin.  In Section \ref{norm} and \ref{counting} 
we show that these decoupling states can
have positive-definite norms and that the corresponding Verma modules are
unitary irreducible representations of Virasoro algebra.  When $q$ is
a root of unity of odd order, in the
semi-classical limit these states give the Bekenstein-Hawking entropy.

\section{Classical Liouville field theory}
\label{sec:classical}
\setcounter{equation}{0}

Here we give a quick review of classical Liouville field theory.
Consider the Liouville action with Euclidean signature \cite{seiberg, ginsparg}
\begin{eqnarray}
\label{euli}
  I_L = \frac{1}{4 \pi} \int d^2 x \sqrt{ \hat{g}} \,
        [ \, \frac{1}{2} \, \hat{g}^{ab} \partial_{a} \phi \, \partial_{b} \phi
         + \frac{ \mu }{2 \gamma^2} e^{\gamma \phi} 
         + \frac{1}{\gamma} R(\hat{g}) \, \phi \, ],
\end{eqnarray}
where $\hat{g}_{ab}$ is the fixed background metric and $\phi$ is the
Liouville field.  The coupling constant $\gamma$ is related to the
cosmological constant by $\gamma^2 = 8 G /l$, and $R$ is the scalar
curvature of the background metric. 

Classically the action (\ref{euli}) defines a conformal field theory
invariant under the Weyl transformation
\begin{align}
  \hat{g}_{ab} \rightarrow e^{2 \rho} \hat{g}_{ab}, ~~
  \gamma \phi \rightarrow \gamma \phi - 2 \rho.
\end{align}
We also define a new field $\varphi = \gamma \phi$, which will be
convenient when we discuss the connection between Liouville
theory and quantum geometry.  In terms of $\varphi$  the action becomes
\begin{align}
    I_L =  \frac{1}{8 \pi \gamma^2} \int d^2 x \sqrt{\hat{g}} \,
        [ \, \hat{g}^{ab} \partial_{a} \varphi \,
        \partial_{b} \varphi + 4 e^{\varphi} + R(\hat{g}) \, \varphi \, ]. 
\end{align}
Here $\mu$ is set to be $4$ by a shift in the value of $\varphi$, so
that each solution of the equation of motion corresponds to a
two-dimensional surface with constant Gaussian curvature $K=-1$.  

In terms of the complex coordinates, 
\begin{align}
   z=e^{\tau+ i\sigma}, 
\end{align}
the improved energy-momentum tensor $T$ is given by 
\begin{align}
\label{emtensor}
  T_{zz} = \frac{1}{\gamma^2} (\varphi_{zz} - \frac{1}{2} \,
  \varphi^2_z)
\end{align}
with Fourier modes that satisfy the Virasoro algebra
\begin{align}
\label{c_virasoro}
  i \{L_m, L_n\}_{\rm P.B.} = 
      (m-n) \, L_{m+n} + \frac{c_{\text{cl}}}{12} \, (m^3 -m)
      \delta_{m,-n}
\end{align}
with a classical central charge $c_{\text{cl}}=12/\gamma^2$ (see, for
example \cite{seiberg, ginsparg}).  

\subsection{Classical solutions in Euclidean space}
\label{sec:solutions}

One can choose a local coordinate system such that $\hat{g}_{ab} =
\delta_{ab}$.  The equation of motion for $\varphi$, called the
Liouville equation \cite{liouville}, is then
\begin{align}
\label{eom3}
  \partial_z \partial_{\bz}\,  \varphi = \frac{1}{2} e^{\varphi}.
\end{align}
By the uniformization theorem, each solution of (\ref{eom3}) 
\begin{equation}
\label{solution}
  e^{\vpc(z)} dz d \bar{z} 
  = 4 \frac{\partial A(z) \bar{\partial }B(\bar{z})}
             {[A(z) + B(\bar{z})]^2} dz d \bar{z},
\end{equation}
describes a two-dimensional surface with
constant negative Gaussian curvature $K=-1$ conformally equivalent to a quotient of
the Poincar$\e$ upper half plane $H$ by a discrete subgroup $\Gamma
\in PSL(2,R)$, for some locally defined
(anti-)holomorphic functions $A(z) (B(\bz))$.

 Along a curve $z \rightarrow e^{2 \pi i} z$, $A$ and $B$ transform by an 
$SL(2,R)$ transformation.  Depending on the conjugacy class of the
monodromy of $A(z)$ and $B(\bz)$, there are three classes of local solutions:
elliptic, parabolic and hyperbolic. 
\begin{enumerate}
\item elliptic:~~~~~ $A(z)  \rightarrow (T R_{\theta}\, T^{-1}) A(z)$,
the curve surrounds a conical singularity on the surface, 
\item parabolic: ~~$A(z) \rightarrow (T P_{\lambda} \, T^{-1}) A(z)$,
the curve surrounds a puncture on the surface, 
\item hyperbolic: ~$A(z) \rightarrow (T B_{\epsilon}\, T^{-1}) A(z)$,
the curve surrounds a handle of the surface,
\end{enumerate}
where
\begin{eqnarray*}
  R_{\theta} = 
  \left( \begin{array}{cc}
     \cos \theta & \sin \theta  \\
    -\sin \theta & \cos \theta  
  \end{array} \right), ~
  P_{\lambda} = 
  \left( \begin{array}{cc}
      1 &  \lambda  \\
      0 &  1 
  \end{array} \right), ~
  B_{\epsilon} = 
  \left( \begin{array}{cc}
     e^{\pi \epsilon} & 0  \\
     0 & e^{- \pi \epsilon}  
  \end{array} \right),
\end{eqnarray*}
and $T$ is a $PSL(2,C)$ (M$\rm \ddot{o}$bius) transformation of $A(z)$.
$B(\bz)$ transforms in a similar fashion.

\begin{figure}
  \centerline{\includegraphics[scale=0.95]{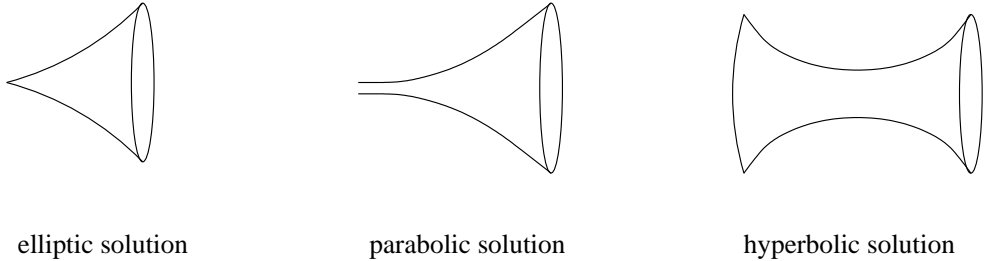}}
  \caption{$\sigma$-independent Euclidean solutions} \label{fig:solution}
\end{figure}

Here are some $\sigma$-independent examples of the classical solutions,
shown in Fig. \ref{fig:solution},
which are :
\begin{enumerate}
\item elliptic solution:
\begin{align}
  e^{\varphi} =  \frac{4 a^2}{(z \bz)^{1-a} [1-(z \bz)^a]^2}, ~~~
  T(z) = \frac{1}{z^2} \Big ( - \frac{a^2}{2 \gamma^2} + \frac{1}{2 \gamma^2}\Big ),
\end{align}
\item parabolic solution:
\begin{align}
  e^{\varphi} = \frac{4}{z \bz \, [\log z \bz]^2}, ~~~
  T(z) = \frac{1}{z^2} \frac{1}{2 \gamma^2},
\end{align}
\item hyperbolic solution:
\begin{align}
  e^{\varphi} =  \frac{4 m^2}{z \bz \,  \big [\sin \big (\frac{m}{2}
  \log z \bz \big )\big ]^2}, ~~~
  T(z) = \frac{1}{z^2} \Big ( \frac{m^2}{2 \gamma^2} + \frac{1}{2 \gamma^2}\Big ).
\end{align}
\end{enumerate}

\subsection{The classical $SL(2,C)$ symmetry}
\label{classical-symmetry}

Consider the system defined on a flat cylinder, where $\sigma \in [0, 2
\pi)$ parametrizes space and $\tau$ parametrizes imaginary time.  
Define light-cone coordinates $x^{\pm} = \sigma \mp i \tau$. 

A Liouville solution with periodic boundary condition can be rewritten in terms of the
chiral fields $\chi_i(x^+)$, $\xi_i(x^-)$, $i=1,2$, with conformal
weights $(-1/2, 0)$ and $(0, -1/2)$:
\begin{equation}
\label{chi}
  e^{- \varphi} = \frac{1}{4} (\chi_2 \xi_1 + \chi_1 \xi_2)^2, 
\end{equation}
where $\chi_i(x^+)$, $\xi_i(x^-)$ are two pairs of real solutions of the 
Schr{\"o}dinger equations
\begin{align}
\label{schr}
  (\,  \partial_+^{\, 2} - \frac{\gamma^2}{2} T_{++}\, ) \, \chi_i &= 0
                ~~~~~ i=1,2  \non \\
  (\,  \partial_-^{\, 2} -  \frac{\gamma^2}{2} T_{--}\, ) \, \xi_i &=
  0, 
\end{align}
with unit Wronskians $\chi_1 \chi'_2  - \chi'_1 \chi_2 = 1$, 
$\xi_1 \xi'_2  - \xi'_1 \xi_2 = 1$.  The real periodic potentials 
$T_{\pm \pm} = \frac{1}{\gamma^2} ( - \varphi_{\pm \pm} + \frac{1}{2}
\varphi^2_{\pm})$ satisfy 
$ T_{\pm \pm}(\sigma + 2 \pi) =  T_{\pm \pm}(\sigma)$. 

Note that a constant M{\"o}bius transformation
\begin{align}
  \left( \begin{array}{cc}
     \chi_1 \\
     \chi_2  
  \end{array} \right) \rightarrow
  \left( \begin{array}{cc}
     a & b  \\
     c & d  
  \end{array} \right)
  \left( \begin{array}{cc}
     \chi_1 \\
     \chi_2  
  \end{array} \right),~~~
  \left( \begin{array}{cc}
     \xi_1 \\
     - \xi_2  
  \end{array} \right) \rightarrow -
  \left( \begin{array}{cc}
     a & b  \\
     c & d  
  \end{array} \right)
  \left( \begin{array}{cc}
     \xi_1 \\
     -\xi_2  
  \end{array} \right)
\end{align} 
preserves both 
$e^{- \varphi} = \frac{1}{4} (\chi_2 \xi_1 + \chi_1 \xi_2)^2$
and the unit Wronskians.  For $2j$ a positive integer, we have primary
fields $e^{- j \varphi}$ with weights $(-j,-j)$ expressed as
\begin{eqnarray}
\label{jspin}
  e^{ - j \varphi} &=& \Big ( \frac{1}{4} \Big )^2 
                            (\chi_2 \xi_1 + \chi_1 \xi_2)^{2j}  \non \\  
                       &=& \Big ( \frac{1}{4} \Big )^2 
                           \sum_{m=-j}^{j} (-1)^{j-m} \, \psi^j_m(x^+)
                            \, \bar{\psi}^j_{-m}(x^-),
\end{eqnarray}
where the fields
\begin{align}
  \psi^j_m(x^+) = \sqrt{
  \left( \begin{array}{cc}
     2j \\ j-m 
  \end{array} \right)}~ \chi_1^{j-m} \chi_2^{j+m}, ~~
  \bpsi^j_{-m}(x^-) = \sqrt{
  \left( \begin{array}{cc}
     2j \\ j-m 
  \end{array} \right)}~ \xi_1^{j+m} (-\xi_2)^{j-m}
\end{align}
transform under $SL(2,C)$ transformations like the
spin-$j$ representation with finite dimension $2j+1$.  
For $2j$ a negative integer, on the other hand, the
decomposition into chiral fields is infinite.  These fields also
form $SL(2,C)$ representation. 

\section{Canonical quantization}
\label{quantization}
\setcounter{equation}{0}

\subsection{B{\"a}cklund transformation}
\label{sec:backlund}

For canonical quantization, we may work at imaginary time $\tau =0$ without loss of generality.
Because the potential $T$ of the Schr{\"o}dinger equation
(\ref{schr}) is periodic, $T(\sigma+ 2 \pi) = T (\sigma)$, the solution has periodicity 
\begin{align}
  \chi_i(\sigma+ 2 \pi) = M^j_i \, \chi_j (\sigma),
\end{align}
where $M^j_i$ is the monodromy matrix of the Schr{\"o}dinger equation,
which is related to the monodromy property of the classical solution.
For the hyperbolic and elliptic solutions, we can use an $SL(2,C)$
transformation to diagonalize $M^j_i$ so that
$\chi_i$ (and $\xi_i$) becomes periodic up to a multiplicative constant 
\begin{align}
\label{peri}
  \chi_1(\sigma + 2 \pi) = e^{\gamma \pi p_0^{(1)}} \chi_1(\sigma) ,~~~
  \chi_2(\sigma + 2 \pi) = e^{\gamma \pi p_0^{(2)}} \chi_2(\sigma),  
\end{align}
where $p_0^{(i)}$ are real for hyperbolic solution and imaginary for
elliptic solution.  We will
not consider the parabolic solution in this paper.
A priori $\chi_i$, $\xi_i$  do not need to
be real, as long as the resulting $\varphi$ field is real.

The chiral fields $\chi_{i}$ can be written in terms of bosonic fields $q^{(i)}$  as 
\begin{equation}
\label{boson}
 \chi_{i} = \exp{(\frac{\gamma}{2} \, q^{(i)})}.
\end{equation}
With the chosen periodicity (\ref{peri}) the fields $q^{(i)}$ and
$\bq^{(i)}$ can be expanded in Fourier series:
\begin{align}
  q^{(i)}(\sigma) &= q^{(i)}_0 + p^{(i)}_0 \sigma + 
                 i \sum_{m \neq 0} a^{(i)}_m e^{-i m \sigma}/m, 
                 ~~~ i=1,2 ,
\end{align}
and it can be shown that the Fourier modes \cite{curtright2,gn1, gn2,
  gn3b, gn3, ger1} satisfy the Poisson brackets
\begin{align}
  \{ a^{(i)}_n, a^{(i)}_m \}_{\rm P.B.} =  - i n \, \delta_{n,-m}.
\end{align}
Assuming in addition that $\{q^{(i)}_0, p^{(i)}_0\}_{\rm P.B.}
=1$, the Liouville field $\varphi(\sigma)$ is thus related to the free fields
$q^{(i)}(\sigma)$ and the anti-chiral counterparts
$\bq^{(i)}(\sigma)$ 
through a classical canonical transformation (a B{\"a}cklund transformation).

The stress energy tensor $T_{++}$ can be expressed in terms of either
set of the free fields \cite{gn3b, gn3} as
\begin{align}
\label{tensor}
  T_{++} 
   = \frac{1}{\gamma} \, [q^{(1)} {}'' + \frac{\gamma}{2} (q^{(1)}{}')^2 ]
   = \frac{1}{\gamma} \, [q^{(2)} {}'' + \frac{\gamma}{2} (q^{(2)}{}')^2].
\end{align}
The conformal generators can be written in either set of the
Fourier modes, and give a Poisson-bracket realization of the Virasoro algebra (\ref{c_virasoro}).

\subsection{Canonical quantization}
\label{sec:canonical}

After the B{\"a}cklund transformation, since
$q^{(i)}(\sigma)$ and $\bq^{(i)}(\sigma)$ are free fields,
the quantization is straightforward.  All the complications of
the interacting theory are in the B{\"a}cklund transformation.  

Classically the two free fields $q^{(1)}(\sigma)$
(or $\bq^{(1)}(\sigma)$) 
and $q^{(2)}(\sigma)$ (or $\bq^{(2)}(\sigma)$) are not independent.
We could express one of the free field in terms of the other and proceed with the
canonical quantization in terms of one free field as in
\cite{curtright1,curtright2}.   Instead, we will follow the
canonical quantization procedure of Gervais and Neveu
\cite{gn1,gn2,gn3b,gn3,gn4,gn5,gn6,gn7,ger1,ger2,ger3}, where the two free fields remain
symmetric in the quantization, so that the quantum algebra structure 
becomes clear.

The basic idea of \cite{gn1,gn2,gn3b,gn3,gn4,gn5,gn6,gn7,ger1,ger2,ger3} is
quite natural.  Classically  we have the relation (\ref{tensor}), where the
description in terms of two sets of
fields are equivalent.
Now impose canonical quantization conditions such that the two
sets of the free fields remain symmetric:
\begin{align}  
  [q^{(i)}_0, p^{(i)}_0] =i, ~~~&~~~
  [a^{(i)}_n, a^{(i)}_m]  = n \, \delta_{n,-m}, \\
  N^{(1)} \, [q^{(1)} {}'' + \frac{\gamma}{2} (q^{(1)}{}')^2 ]
    &= N^{(2)} \,[q^{(2)} {}'' + \frac{\gamma}{2} (q^{(2)}{}')^2]
  (\bq^{(2)}{}')^2], 
\label{equal}\\
  p_0 &\equiv p^{(1)}_0 = - p^{(2)}_0. \label{momenta}
\end{align}
In (\ref{equal}) $N^{(i)}$ define normal
orderings for the two sets of fields.  Condition
(\ref{momenta}) ensures that the Wronskians will not change when 
$\sigma \rightarrow \sigma + 2 \pi$. 

The Virasoro generators can then be written in terms of either set of the
normal-ordered creation-annihilation operators \cite{gn3b, gn3}
\begin{eqnarray}
  T_0 &=& N^{(i)} \, (\frac{1}{2} \, p_0^{(i) \, 2} + \sum^{\infty}_{n=1} 
                   a_{-n}^{(i)} a_n^{(i)}), \nonumber  \\
  T_m &=& N^{(i)} \, ( p_0^{(i)} a_m^{(i)}  - \frac{i}{\gamma} \,  m a_m^{(i)} 
           + \frac{1}{2} \sum_{n \neq 0,m} a_{m-n}^{(i)} \, a_n^{(i)}),
  ~~ m \neq 0,
\end{eqnarray}
and satisfy 
\begin{eqnarray}
  [ T_m, T_n ]    
  =  (m-n) (T_{m+n} - \frac{1}{24} \delta_{m,-n}) \,  
      + \frac{1}{12} (1+ \frac{12}{\gamma^2}\, ) \, m^3 \delta_{m,-n}
\end{eqnarray}
with quantum central charge  
\begin{align}
\label{quantum_c}
  c=1+ \frac{12}{\gamma^2}.  
\end{align}
Mapped onto the complex plane, the operators
\begin{eqnarray}
  L_0 = T_0 + \frac{1}{2 \gamma^2}, ~~~~~~ L_m = T_m,
\end{eqnarray}
satisfy the standard Virasoro algebra.

\subsection{Quantum coupling constant}
\label{renorm}

After canonical quantization, the conformal weight of the vertex operator $e^{\alpha \varphi}$
is shifted from the classical value to
\begin{align}
  \Delta = \bDelta = - \frac{\gamma^2}{2} \, (\alpha - \frac{1}{\gamma^2})^2 +
  \frac{1}{2 \gamma^2}.
\end{align}
It is convenient to define a quantum coupling constant $\tgamma$ \cite{gn4,gn5}.
The quantum operator corresponding to the metric is no longer
$e^{\varphi}$ but is renormalized to $e^{ \nu \varphi}$,  with
conformal weight $(1,1)$, such that $e^{\nu \varphi} dz d \bz$ remains
conformally invariant.  The quantum coupling
constant is defined by $\tilde{\gamma} = \nu \gamma$, and satisfies the 
equation
\begin{equation}
\label{tgamma}
  \frac{\tilde{\gamma}}{ \gamma} - \frac{1}{2} \tilde{\gamma}^2 =1,
\end{equation}
which has two solutions:
\begin{equation}
  \tilde{\gamma}_{\pm} = \frac{1 \pm \sqrt{1 - 2 \gamma^2}}{\gamma}.
\end{equation}
It is this quantum coupling constant $\tgamma$ which will appear in
the deformation parameters in the quantum algebra.

In terms of the quantum coupling constant $\tgamma$, the central
charge (\ref{quantum_c}) now becomes
\begin{equation}
\label{charge}
  c= 1+ 3 \, (\frac{2}{\tilde{\gamma}} + \tilde{\gamma})^2,
\end{equation}
or equivalently
\begin{equation}
  \tilde{\gamma}^2 = \frac{c-13 \pm \sqrt{(c-1)(c-25)}}{6}. 
\end{equation} 
The renormalization of the coupling constant is the same as in \cite{dhoker}.  

In this paper we are interested in the region $c>25$, the so-called
``weak-coupling region'', which corresponds to the semi-classical region
$l \gg G$.  In this region, both $\gamma$ and $\tgamma$ are real.  In
the following we will write $\tgamma_-$ as
$\tgamma$ and $\tgamma_+$ as $2/\tgamma$. 

\subsection{The nonexistence of $SL(2,C)$ invariant vacuum}
\label{vacuum}

Define the operator \cite{gn4}
\begin{align}
  \w = i \, \frac{2}{\tilde{\gamma}} \, p_0,
\end{align}
where the zero-mode $p_0$ is defined in (\ref{momenta}).  For $c>25$ and
$\tilde{\gamma}$ real, since the spectrum $\up_0$ of $p_0$ 
is purely imaginary in the elliptic sector, the spectrum $\uw$ of
$\varpi$ is real.   In the hyperbolic sector, on the other hand, the spectrum of $\varpi$ is purely
imaginary.

Introduce a basis of zero mode eigenstates $| \uw \ket$ such that
\begin{align}
  \w \,| \uw ,0 \,\ket = \uw \,| \uw,0 \, \ket, 
   ~~~~~~ a^{(i)}_n | \uw, 0 \,\ket =0, ~~~ n>0, ~i=1,2 .       
\end{align}
These are highest-weight states, satisfying \cite{gn4}
\begin{align}
\label{highest}
  L_0 \, | \uw , 0 \, \ket  = \Big [ \frac{1}{2 \gamma^2} \,  
         +\frac{1}{2} \, \up^{(1) \, 2}_0  \Big] \, | \uw  , 0\, \ket 
      &=  \Big[ \frac{1}{2 \gamma^2} \,
         +\frac{1}{2} \, \up^{(2) \, 2}_0  \Big] \, | \uw  , 0 \, \ket,  \non \\
  L^{(1)}_n |  \uw , 0\, \ket = L^{(2)}_n |  \uw  , 0\, \ket &=0, \hspace{2cm}  n>0. 
\end{align}
The Hilbert space is a direct sum of Verma modules (not necessarily irreducible) which are obtained
by applying either $L^{(1)}_{-n}$ or  $L^{(2)}_{-n}$ for $n>0$ on the highest-weight states labeled by $\uw$.
The two chiral Verma modules generated by $p^{(1)}_0$ and $p^{(2)}_0$
coincide, since the highest weights only depends upon $\up^{(1)\, 2}_0 = \up^{(2)\, 2}_0$.  

The ground state structure of quantum Liouville theory is rather
unconventional.  The canonical quantization condition (\ref{equal})
can be solved perturbatively \cite{gn3}, yet the (quantum) series
expansion for one set of oscillators in terms of the other has
manifest poles and breaks down at certain values of zero modes
\begin{equation}
  \up_0 = im / \gamma, ~~~~ m \in \mathbf{Z}  \mathnormal \backslash \{ 0 \}.
\end{equation} 
Classically for these values of $\up_0$, the Liouville field $\varphi$
cannot be real.  The value $\up_0 = i /\gamma$ corresponding to $ \Delta
= 0$ is included in this set of zero modes, which means that the
$SL(2,C)$ invariant vacuum is not included in the spectrum.  
We can see this explicitly \cite{schnittger} from the fact that the
$L_{-1}$ (translation) operator 
acting on a ground state 
\begin{equation}
   L^{(i)}_{-1}  |\,  \uw \, \rangle 
   = \Big (p_0^{(i)} a_{-1}^{(i)} + \frac{i}{\gamma} \, a_{-1}^{(i)}
           + \frac{1}{2} \sum_{n \neq 0, -1} a_{-1-n}^{(i)} \,
   a_n^{(i)} \Big ) \, | \, \uw \, \ket = 0
\end{equation}
implies that $\up_0^{(1)} = \up_0^{(2)} = -i / \gamma$, which 
contradicts with the condition that $p_0^{(1)} = - p_0^{(2)}$.  

D'Hoker and Jackiw \cite{jackiw} argue from the quantum equation of
motion 
\begin{align}
   \nabla \varphi = 2 e^{\eta \varphi}
\end{align}
that no translationally invariant normalizable vacuum $|0 \ket$ exists in
Liouville theory, since that would imply that 
\begin{align}
  \bra \, 0 \, | \,2 e^{\eta \varphi} \,| \, 0 \, \ket 
   = \bra  \, 0 \,| \,\nabla \varphi \,| \, 0 \,\ket =0,
\end{align}
violating the formal positivity of the exponential.  We will see
in Section \ref{ward identity} that the non-existence of
$SL(2,C)$-invariant vacuum in the spectrum has important
consequences in the whole Hilbert space structure of the quantum theory.

\pagebreak

\subsection{Construction of Hartle-Hawking states}
\label{HH-states}

\begin{wrapfigure}[11]{l}{1.9in}
  \begin{center}
    \epsfig{file=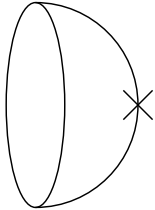, height=80pt}
    \caption{Operator insertion}
  \end{center}
\end{wrapfigure}

In the framework of standard conformal field theory, a basic assumption is that
all highest-weight states are generated by applying a primary field to the 
$SL(2,C)$-invariant vacuum $| \,0 \ket \, $.  As we already discussed
above, the situation is not
standard in Liouville theory because of the absence of the
$SL(2,C)$-invariant 
vacuum.  The nonexistence of such a vacuum in the spectrum means that the standard map
from an operator $\CO$ to the state $\CO(z=0) | 0 \ket$ cannot be used
here.  Instead, we can create a Hartle-Hawking state by
performing a path integral on a disk $D$. To evaluate such a path integral,
the boundary conditions for the field $\varphi$ must be specified
on the boundary of the disk.
Insertion of the operator $\CO$ on the disk $D$ gives
the wavefunction $\Psi_\CO$ for the operator $\CO$
\begin{align}
  \Psi_{\CO} \big[\varphi_b \big]
=\int_{\varphi|_{\partial D}=\varphi_b} [d \varphi]\ {\rm e}^{-I[\varphi]}\,\CO\ .
\end{align}
The insertion of the vertex operator $\CO = e^{\alpha \varphi}$
creates a state with purely imaginary zero mode $\up_0 = i (\gamma
\alpha - 1/ \gamma)$, corresponding to
a highest-weight state in the elliptic sector. 

\subsection{Spectrum of the Hilbert space}
\label{hilbert1}

The spectrum of the theory thus includes different sectors depending
on the values of the zero modes $\up_0$:
\begin{align}
\label{spectrum}
  \text{hyperbolic sector: ~~$\up_0$ real}, 
   ~~~~~~~~~~~~~\Delta = 1/ (2 \gamma^2) + \, \up^2_0\, / 2 \, > \,
   1/(2 \gamma^2), \non \\
  \text{elliptic sector:  ~~~~~~~~$\up_0$ imaginary}, ~~~ \Delta = 1/(2 \gamma^2) +
   \, \up^2_0\, /2\, <  \, 1/(2 \gamma^2).
\end{align}

The distinction between these two sectors can be made clear
\cite{seiberg} in 
the mini-superspace approximation of the theory, where Liouville
theory is described by a quantum mechanics problem of
$\sigma$-independent $\phi_0$.
The states in the hyperbolic sector are labeled by the real continuous
parameter $\up_0$, and the wave function is normalizable
in the limit $\phi_0 \rightarrow - \infty$, when interaction term vanishes.  The states
that correspond to local vertex operator insertions are in the
elliptic sector of the theory and lead to eigenfunctions of the
Hamiltonian which diverge as $\phi_0 \rightarrow - \infty$ because of
the imaginary value of $\up_0$ .  Thus
these Hartle-Hawking states are called nonnormalizable states.

The general solutions of (2+1)-dimensional gravity with $\Lambda<0$ can be
classified in terms of the spectrum of Liouville theory. 
The Fefferman-Graham expansion \cite{fefferman} of the metric which
solves the Euclidean Einstein's equation is completely defined by the 
geometry on the boundary \cite{banados2,skenderis,rs2}.  Choosing
the asymptotic geometry to be an infinite cylinder, the complete
expression of a locally AdS metric is
\begin{align}
\label{expansion}
    ds^2 = 4 \frac{G}{l} (L d\omega^2 + \bar{L} d\bar{\omega}^2) 
         + (e^{2 \rho} + 16 \frac{G^2}{l^2} L \bar{L} e^{-2 \rho}) d \omega d \bar{\omega}
         +l^2 d \rho^2,
\end{align}
where $\{\omega, \bomega, \rho \}$ are coordinates such that the
boundary is located at $e^{\rho} \rightarrow \infty$, and  $\omega$,
$\bomega$ are complex coordinates on the boundary.  

When $L$ and $\bL$ have constant values $L_c$ and $\bL_c$, we can
parametrize them as 
\begin{align}
  L_c= \frac{1}{2} (Ml + J) , ~~~~~~ \bL_c = \frac{1}{2} (Ml -J).
\end{align}
For $Ml > |J|$, the metric (\ref{expansion}) is globally isometric to the Euclidean (2+1)-dimensional
black hole of mass $M$ and angular momentum $J$.  For $J=0$ and $M
=-1/8G$ the metric reduces to Euclidean anti-de Sitter space.  When $Ml < |J|$, the metric can be
described as a conical singularity, which has an ADM mass lying between the anti-de Sitter value
of -1/8G and the extremal BTZ black hole value of zero. 

The functions $L$ and $\bL$ in the exact solution are given by the
energy-momentum tensor in the boundary Liouville field theory \cite{skenderis,rs2}. 
Comparing the spectrum (\ref{spectrum}) with the values of $L_c$ and
$\bL_c$, we see that the hyperbolic sector of Liouville field theory
correspond to black hole solutions, while the elliptic sector give
solutions which behave like conical singularities.

\section{The underlying quantum algebra structure for Liouville theory}
\label{sec:group}
\setcounter{equation}{0}

We have seen from (\ref{jspin}) that under $SL(2,C)$ 
transformations the chiral fields $\psi^j_m(\sigma)$ transforms
classically in the
spin $j$ representation with finite dimension $2j+1$, and that $e^{-j
  \varphi}$ is a group invariant.  Gervais and Neveu 
\cite{gn4,gn5,gn6,gn7,ger1,ger2,ger3} have shown that this group 
structure is replaced by a quantum algebra after canonical quantization.

\subsection{The exchange algebra and fusion of chiral vertex operators}
\label{sec:exchange}

Define normal-ordered chiral vertex operators 
\begin{align}
\label{chiral}
  \psi_i = \, d_i(\varpi) N^{(i)}(\exp{(\frac{\tilde{\gamma}}{2} \, q^{(i)})})
\end{align}
as normalized solutions of the quantum version of eqn. (\ref{schr}),
where $d_i(\varpi)$ is normalization factor depending only on zero-modes. 
The $\psi_i$ fields satisfy the exchange algebra 
\begin{equation}
\label{exchange} 
 \psi_i(\sigma) \psi_j(\sigma') 
  = \sum_{k=1,2; \, l=1,2} S^{kl}_{ij} \, (\varpi, \sigma - \sigma')\,
          \psi_k(\sigma') \, \psi_l(\sigma).
\end{equation}
It is shown in  \cite{gn4} that the above $S^{kl}_{ij}$ satisfiesthe
Yang-Baxter equations \cite{yang,baxter72,baxter82}
\begin{eqnarray}
    \sum_{\rho, \lambda, \mu} 
    S^{\lambda \mu}_{jk}(\varpi + \varepsilon_i, \sigma_2 - \sigma_3) \,
    S^{l \rho}_{i \lambda}(\varpi, \sigma_1 - \sigma_3) \,
    S^{mn}_{\rho \mu}(\varpi + \varepsilon_l, \sigma_1 - \sigma_2)
     \nonumber \\
  = \sum_{\rho, \lambda, \mu} 
    S^{\mu \lambda}_{ij}(\varpi, \sigma_1 - \sigma_2) \,
    S^{\rho n}_{\lambda k}(\varpi + \varepsilon_{\mu}, \sigma_1 - \sigma_3) \,
    S^{lm}_{\mu \rho}(\varpi, \sigma_2 - \sigma_3), \\
  \varepsilon_2 = - \varepsilon_1 = 1, \nonumber
\end{eqnarray}
due to the associativity of the products of three $\psi$ fields.
However they depend upon the zero modes $\varpi$, which can be shifted by
the $\psi$ fields since 
\begin{equation}
   \varpi \, e^{\frac{\tgamma}{2} q^{(i)}_0} = e^{\frac{\tgamma}{2}
   q^{(i)}_0} \, (\varpi \pm 1) \, .
\end{equation}

Gervais \cite{ger2} showed that by taking operator product of $\psi_i, i=1,2$,
one can generate chiral fields
\begin{equation}
  \psi^{\mu, \nu} \sim
  N^{(1)} (e^{\mu \frac{\tilde{\gamma}}{2} q^{(1)}}) N^{(2)} (e^{\nu \frac{\tilde{\gamma}}{2} q^{(2)}}),
\end{equation}
with  integer $\mu$, $\nu$.
It is convenient to adopt the notation:
\begin{equation}
  \psi^{(j)}_m \equiv \psi^{j-m, j+m};~~~ \mu + \nu = 2j, ~\nu - \mu =2m,
\end{equation}
with $\psi^{(1/2)}_{-1/2}= \psi_1$ and $\psi^{(1/2)}_{1/2}= \psi_2$.
The positive half-integer $j$ determines the conformal weight of $\psi^{(j)}_m$:
\begin{equation}
  \Delta_j = -j - \frac{\tilde{\gamma}^2}{2} j(j+1).
\end{equation}

The operators $\psi^{(j)}_m$ are closed under OPE and braiding,
but the fusion coefficients and $R$-matrix elements depend on
the zero modes $\w$ and thus do not commute with the $\psi^{(j)}_m$. 
The quantum-group structure can be exhibited more explicitly, once one
changes
to another basis of Hermitian chiral fields  \cite{ger2}
\begin{equation}
  \xi^{(j)}_M(\sigma) 
    =\sum_{-j \leq m \leq j} \,a^m_M( j, \varpi )\,
    \psi^{(j)}_m(\sigma), ~~~ -j \leq M \leq j.
\end{equation}
It was shown in \cite{ger2} that the
exchange algebra of these operators is
\begin{equation}
  \xi^{(j)}_{M}(\sigma) \, \xi^{(j')}_{M'}(\sigma') 
  = \sum_{-j \leq N \leq j; -j' \leq N' \leq j'} (j,j')^{N'N}_{M M'} 
    \xi^{(j')}_{N'}(\sigma') \, \xi^{(j)}_{N}(\sigma),
\end{equation}
where $(j,j')^{N'N}_{M M'}$ is coefficient of the universal $R$-matrix
of quantum algebra $U_h(sl_2)$
\begin{eqnarray}
\label{r1}
  R = e^{h H \otimes H/2} 
      \sum_{n=0}^{\infty} \frac{ q^{n(n+1)/2}\, 
      (1- q^{-2})^n}{[ n ] !} \, J_+^n \otimes J_-^n,
\end{eqnarray}
 with 
\begin{align}
  h= i \pi \tgamma^2/2,
\end{align}
where $q=e^{-h}$ for $0 < \sigma< \sigma' < \pi$ and
$q=e^{h}$ for $0 < \sigma' < \sigma < \pi $, and
\begin{align}
\label{deform_n}
  [n] = \frac{q^n- q^{-n}}{q- q^{-1}}.
\end{align}
$H$, $J_+$ and $J_-$ are the generators of the the formal deformation 
algebra $U_h(sl_2)$ of the univeral enveloping algebra $U(sl_2)$.   
The algebra $U_h(sl_2)$ (see \cite{klimyk, chari, kassel} and the references therein)
is an $h$-adic algebra defined over the ring
$\C[[h]]$ of formal power series with the defining relations\footnote{Note that
  in Gervais and Neveu's work the definition of $U_h(sl_2)$ is
  slightly different, such that their generator $J_3$ is twice the
  generator $H$ defined here.}
\begin{align}
\label{algebra}
  [H, J_{\pm}] = \pm 2 J_{\pm}, ~~~~~
  [J_+, J_-] = [H] \equiv \frac{e^{h H} - e^{-h H}}{e^h -e^{-h}}.
\end{align} 
It admits a unique $h$-adic Hopf algebra structure and is
quasitriangular (see Appendix \ref{definition} for some 
basic definitions regarding quantum algebras).

The short-distance operator-product expansion of the $\xi$ fields is of the 
form \cite{ger2}
\begin{align}
\label{fusion}
  \xi^{(j_1)}_{M_1}(\sigma) \,& \xi^{(j_2)}_{M_2}(\sigma') 
  = \sum^{j_1 + j_2}_{j = | j_1 -j_2 |} 
    \Big \{ (1-e^{-i(\sigma -\sigma')})^{\Delta(j) - \Delta(j_1) - \Delta(j_2)}
    \times  \nonumber \\
  & \times \, (j_1, M_1; j_2, M_2 | j_1, j_2; j, M_1 + M_2)
       \Big ( \xi^{(j)}_{\, M_1 + M_2} ( \sigma ) + \rm descendants 
              \Big ) \Big \},
\end{align}
where $(j_1, M_1; j_2, M_2 | j_1, j_2; j, M_1 + M_2)$ are the 
Clebsh-Gordan coefficients of $U_{q}(sl_2)$,
and $\Delta(j) = -j - \frac{\tilde{\gamma}^2}{2} j(j+1)$ is the 
conformal weight of $\xi^{(j)}_M$, assuming that there is no maxium vaule of $j$.

\subsection{Quantum algebras $U_q(sl_2)$ and 
representations with $q$ as roots of unity}
\label{usl2}

Let $q$ be a fixed complex number such that $q \neq 0$ and $q^2 \neq
1$.  Denote by $U_q(sl_2)$ the quantum algebra \cite{klimyk, chari, kassel}
over $\C$ with four generators $J_+$, $J_-$, $K$, $K^{-1}$ satisfying the
defining relations
\begin{align}
\label{commutation}
  K K^{-1}= K^{-1}K =1, ~~[J_+, J_-] = \frac{K- K^{-1}}{q-q^{-1}}, 
   ~~ K J_{\pm} K^{-1} = q^{\pm 2} J_{\pm},
\end{align}
on which exists a unique Hopf algebra structure with comultiplication
$\Delta$, counit $\varepsilon$ and antipode $S$ such that 
\begin{align}
  \Delta (J_+) = J_+ \otimes K + 1 \otimes J_+, ~
  \Delta (J_-) &= J_- \otimes 1 + K^{-1} \otimes J_-, ~
  \Delta(K) = K \otimes K, \nonumber \\
  \varepsilon (K) = 1, ~~~&~~~ \varepsilon (J_{\pm}) =0, \nonumber \\
  S(K) = K^{-1}, ~~~~S(J_+) &= -J_+ K^{-1}, ~~~~S(J_-) = - K J_-.
\end{align}
Note that the Hopf algebras $U_q(sl_2)$ and $U_{q^{-1}}(sl_2)$ are isomorphic.

The quantum algebra $U_q(sl_2)$ does not admit a universal $R$-matrix since
it is not quasitriangular, but it is a braided Hopf algebra
\cite{reshetikhin} such that the condition of quasitriangularity is
generalized to the existence of an automorphism of $U_q(sl_2) \otimes
U_q(sl_2)$.  Because of an injective homomorphism 
$i: U_q(sl_2) \rightarrow U_h(sl_2)$ such that
\begin{align}
  i(K) = e^{h H} , ~~~~ i(K^{-1}) = e^{-h H}, ~~~~ i(q) = e^h,
\end{align}
the universal $R$-matrix (\ref{r1}) of the quasitriangular algebra
$U_h(sl_2)$ gives the exchange algebra of two representations of
$U_q(sl_2)$ of the type that we shall discuss below.  Equations
(\ref{r1}) and (\ref{fusion}) show the quantum
algebra structure of the chiral fields $\xi^{(j)}_{M}(\sigma)$, with
the deformation parameter 
\begin{align}
  q= e^{i \pi \tgamma^2/2}.
\end{align}

For real positive $q$ the representation theory of the quantum algebra
is essentially the same as that of the corresponding Lie algebra 
\cite{klimyk, chari, kassel}. 
However, the special case when $q$ is a 
root of unity, $q= \exp(\frac{i2\pi}{p})$, $p \in \Z$, is
different.  Define
\begin{equation}
   \begin{cases}
    \, p'= p &~~~~\text{for odd $p$} \\ 
    \, p'= p/2 &~~~~\text{for even $p$}  .
   \end{cases}
\end{equation} 
Then all of the irreducible representations have finite dimension of
at most $p'$.

At roots of unity, the quantum Casimir invariant 
\begin{equation}
  C_q  = J_+ J_- + \frac{K q^{-1} + K^{-1} q}{(q- q^{-1})^2}
       = J_- J_+ + \frac{K q + K^{-1} q^{-1}}{(q- q^{-1})^2}
\end{equation}
is no longer the only invariants of $U_{q}(sl_2)$.  The center of the algebra is
generated by 
\begin{equation}
  C_q, ~~ (J_+)^{p'}, ~~ (J_-)^{p'}, ~~(K)^{p'}, ~~(K)^{-p'}. 
\end{equation}
These invariants are dependent, and there exists a polynomial relation
among them.  Hence irreducible representations of $U_{q}(sl_2)$ have three
independent labels, which we may take to be the eigenvalues of $(J_+)^{p'}$,
$(J_-)^{p'}$, $(K)^{p'}$ with the Casimir invariant $C_q$
determined by them.  

Let $T$ denote an irreducible representation of  
$U_q(sl_2)$ on a vector space $V$.  Concerning the values of 
the scalars $T(J_+)^{p'}$, $T(J_-)^{p'}$, when $q$ is a root of unity, 
the irreducible representations
of $U_q(sl_2)$ fall into three classes \cite{klimyk, chari, kassel}: nilpotent, cyclic and
semicyclic representations.  In this paper we will only discuss the
discrete nilpotent representations $T_{\omega j}$, defined as follows
\cite{klimyk}:

Let $j$ be an nonnegative integer or half-integer and let
$\omega \in  \{ +1, -1\}$.  Let $V_j$ be a $(2j+1)$-dimensional
complex vector space with basis $\be_M$, $M=-j,-j+1,...,j$.  For
notational convenience, set $\be_{j+1}= \be_{-j-1}=0$.  Define operators
$T_{\omega j}(J_+)$, $T_{\omega j}(J_-)$, $T_{\omega j}(K)$ acting on
$V_j$ by
\begin{align}
\label{Twj}
  T_{\omega j}(J_+) \be_M \, &=\, \sqrt{[j-M][j+M+1]} \,
  \be_{M+1},\non \\
  T_{\omega j}(J_-) \be_M \, &= \, \omega \sqrt{[j+M][j-M+1]} \, \be_{M-1}, \nonumber \\
  T_{\omega j}(K) \be_M &= \omega \, q^{2M} \be_M,
\end{align}  
with $[n]$ defined as in eqn. (\ref{deform_n}).
It can be shown that these operators satisfy the defining relations of
the algebra, and hence define a representation $T_{\omega j}$ of
the algebra $U_q(sl_2)$ on $V_j$. 

For $q$ a root of unity, the representation $T_{\omega j}$ is irreducible if and only if $2j <
p'$.  The representation $T_{\omega j}$, $\omega \in \{+1, -1 \}$, 
$j= 0, \frac{1}{2}, 1,... \frac{p'-1}{2}$, are pairwise inequivalent
and satisfy the condition $T(J_+^{p'})=T(J_-^{p'})=0$.

\section{Hartle-Hawking states}
\label{construction}
\setcounter{equation}{0}

\subsection{Vertex operator as quantum group invariant}
\label{operator}

In Section \ref{sec:group} we discussed the quantum algebra structure of
the chiral fields $\xi^{(j)}_{M}(\sigma)$.  
The discussion equally
applies to the chiral fields defined with the other solution of
eqn. (\ref{tgamma}), with $\tgamma$ replaced by $2/ \tgamma$.  
The chiral 
fields $\hxi^{(\hj)}_{\hM}(\sigma)$ can be constructed, and
exhibit similar quantum algebra structure with the deformation
parameter  
\begin{align}
  \hq = \exp(i 2 \pi / \tgamma^2).
\end{align}
Consider the fields $\xi^{(j)}_M$ and $\hxi^{(\hj)}_\hM$ 
as representations of $U_q(sl_2)$ and $U_{\hq}(sl_2)$ of 
type $T_{\omega j}$,
on linear spaces over $\C$ (will here write 
$T_{\omega j}(J_+)$ simply as $J_+$, etc.):
\begin{align}
    J_+ \, \xi^{(j)}_M
 = \sqrt{[ j - M ] [ j + M +1 ]} 
   \xi^{(j)}_{M+1}, ~~~~~~ 
    &\hJ_+ \, \hxi^{(\hj)}_{\hM}
  = \sqrt{[ \hj - \hM ] [ \hj + \hM +1 ]} 
   \xi^{(\hj)}_{\hM+1},   \non \\ 
   J_- \, \xi^{(j)}_M
 = \omega \sqrt{[ j + M ] [ j - M +1 ]}
  \xi^{(\hj)}_{\hM-1}, ~~~~
   &\hJ_- \, \xi^{(\hj)}_\hM
  = \homega \sqrt{[ j + M ] [ j - M +1 ]}
  \xi^{(\hj)}_{\hM-1}, \non \\ 
   K\, \xi^{(j)}_M = \omega q^{2 M} \,  \xi^{(j)}_M,
   \hspace{8pc} ~~~~~
  &\hK\, \hxi^{(\hj)}_{\hM} = \homega q^{2 \hM} \,  \hxi^{(j)}_\hM. 
\end{align}
Following the discussion by Gervais in \cite{gervais_cmp,local}, for
half-integer $j$ and $\hj$, the chiral field $\xi^{(j \, \hj)}_{M \, \hM}$ is constructed by
fusion of hatted and unhatted chiral fields
\begin{align}
     \xi^{(j)}_{M} \hxi^{(\hj)}_{\hM}
     \sim e^{i \pi (M \hj - \hM j)} \xi^{(j \, \hj)}_{M \, \hM}.
\end{align}
The corresponding quantum algebra structure of the general chiral fields
was observed in \cite{gervais_cmp} to be $U_q(sl_2) \odot
U_{\hq}(sl_2)$, where $\odot$ denotes some
kind of graded tensor product, since the hatted and
unhatted fields commute up to a simple phase when $j$ and $\hj$ are
half-integers.  For continuous spins, however, the commutation becomes
nontrivial, and $j$, $\hj$ lose their individual meanings since $j$ and $\hj$
can no longer be separated \cite{gs, schnittger2}.  Instead, one must
introduce the effective spins \cite{schnittger2}
\begin{align}
  j^e = j +  \frac{2}{ \tgamma^2} \, \hj,  ~~~~~
  M^e = M + \frac{2}{\tgamma^2} \, \hM,
\end{align}
which are appropriate quantum numbers in this case.  The fusion and
braiding may be written in terms of these effective spins, which
were shown to consistently include representations that are
semi-infinite \cite{gs, schnittger2}. 
For half-integer $j$ and $\hj$, it is equivalent to specify either $(j,\hj)$ or $j^e$.  

For half-integer $j$ and $\hj$, the complete quantum algebra action on
the $\xi$ family is given by 
\begin{align}
\label{repre}
    J_+ \, \xi^{(j \, \hj)}_{M \, \hM}
 = \sqrt{[ j - M ] [ j + M +1 ]} 
   \xi^{(j \, \hj)}_{M+1 \, \hM}\,, ~~~~~~ 
    &\hJ_+ \, \xi^{(j \, \hj)}_{M \, \hM}
  = \sqrt{[ \hj - \hM ] [ \hj + \hM +1 ]} 
   \xi^{(j \, \hj)}_{M \, \hM+1}\,,   \non \\ 
   J_- \, \xi^{(j \, \hj)}_{M \, \hM}
 = \omega \sqrt{[ j + M ] [ j - M +1 ]}
   \xi^{(j \, \hj)}_{M-1 \, \hM}\,, ~~~\,
   &\hJ_- \, \xi^{(j \, \hj)}_{M \, \hM}
  = \homega \sqrt{[ \hj + \hM ] [ \hj - \hM +1 ]}
  \xi^{(j \, \hj)}_{M \, \hM-1}\,, \non \\ 
   K\, \xi^{(j \, \hj)}_{M \, \hM} = \omega q^{2 M} \,  \xi^{(j \,
 \hj)}_{M \, \hM}\, ,
   \hspace{8pc} ~~~~~\,
  &\hK\, \xi^{(j \, \hj)}_{M \, \hM} = \homega q^{2 \hM} \, \xi^{(j \,
 \hj)}_{M \, \hM} \, , 
\end{align}
with the restriction that $\omega = \homega$, in order for the two
discriptions in terms of $(j,\hj)$ and $j^e$ to be equivalent.  

The reconstruction of the Liouville field as a quantum invariant was
discussed in \cite{local}, as the quantum analog of the classical
expression (\ref{jspin}):
\begin{align}
\label{construct}
  e^{-(j + \hat{\jmath} \frac{2}{\tgamma^2}) \tgamma \phi} 
 = c_{j \hj} \sum_{M \hM} \frac{1}{[2j]![2 \hj]!} \, (-1)^{j-M + \hj - \hM} 
   &q^{(j-M)(j+M -1)} \hq^{(\hj-\hM)(\hj + \hM -1)} \times \non \\
  &\times \xi^{(j \hj)}_{M \hM}(x_+) \, \bxi^{(j \hj)}_{-M
 -\hM} (x_-),
\end{align}
where $c_{j \hj}$ is a constant depending on $j$, $\hj$ and $\tgamma$.
The quantum algebra structure of the field (\ref{construct}) is of the type 
$U_q(sl_2) \odot U_{\hq}(sl_2) \otimes \overline{U_q(sl_2) \odot
  U_{\hq}(sl_2)}$, and the field has negative
conformal weight
\begin{align}
\label{negativeweight}
  \Delta = \frac{c-1}{24}- \frac{1}{24} 
           [(j + \hj +1) \sqrt{c-1} - (j- \hj) \sqrt{c-25}]^2  .
\end{align}
The anti-chiral fields $ \bxi^{(j \hj)}_{-M -\hM} (x^-)$ has a quantum
group structure similar to eqn. (\ref{repre}).

Define the quantum algebra generators $\J_{\pm}$ and $\cK$ by the coproduct
\begin{align}
  \J_+ = J_+ \otimes \bK + 1 \otimes \bJ_+, ~~~~
  \J_- = J_- \otimes 1 +  K^{-1} \otimes \bJ_-, ~~~~
  \cK = K \otimes \bK, \non \\
  \hat{\J}_+ = \hJ_{\pm} \otimes \bar{\hK} + 1\otimes \bar{\hJ}_{\pm}, ~~~~
  \hat{\J}_- = \hJ_- \otimes 1 +  {\hK}^{-1} \otimes \bar{\hJ}_-, ~~~~
  \hat{\cK} = \hK \otimes \bar{\hK}, 
\end{align}
which naturally satisfies the commutation relations (\ref{commutation}).
For values $\bomega =1$, $\omega = \pm 1$, we then have 
\begin{align}
    \J_{\pm} \, e^{-(j + \hat{\jmath} \frac{2}{\tgamma^2}) \tgamma \phi} = 0, ~~~~~~ 
    \cK \, e^{-(j + \hat{\jmath} \frac{2}{\tgamma^2}) \tgamma \phi}
    = \omega \, e^{-(j + \hat{\jmath} \frac{2}{\tgamma^2}) \tgamma
    \phi}, \non \\
    \hat{\J}_{\pm} \, e^{-(j + \hat{\jmath} \frac{2}{\tgamma^2}) \tgamma \phi} = 0, ~~~~~~ 
    \hat{\cK} \, e^{-(j + \hat{\jmath} \frac{2}{\tgamma^2}) \tgamma \phi}
    = \omega \, e^{-(j + \hat{\jmath} \frac{2}{\tgamma^2}) \tgamma \phi},
\end{align}
implying that the quantized Liouville field is a quantum-group
invariant. 

\subsection{Reducible Verma modules and singular states}
\label{null}

We now construct the highest-weight representations of the Virasoro
algebra corresponding to the local insertions of the vertex operators (\ref{construct}).
Since the holomorphic and antiholomorphic components of the overall 
algebra decouple, representations are obtained by taking their tensor
products.  Denote by $V(c, \Delta)$ and $\bar{V}(c, \bar{\Delta})$ the Verma
modules generated by the sets \{${L_n}$\} and
\{${\bL_n}$\} with central charge $c$ and highest weights $\Delta$ and
$\bar{\Delta}$. The Hilbert space in general is a direct sum of the
tensor products of all conformal dimensions of the theory:
\begin{align}
  \sum_{\Delta, \bar{\Delta}} 
   V(c, \Delta) \otimes \bar{V}(c, \bar{\Delta}).
\end{align}
It may happen that the representations of the Virasoro algebra
comprising the states
\begin{align}
\label{state}
  |\Delta,  \{ \lambda \} \, \ket \equiv L_{-k_1}^{\lambda_1} 
   L_{-k_2}^{\lambda_2}... L_{-k_n}^{\lambda_n} 
  |\Delta \ket,  ~~ \lambda_i >0 
\end{align}
are reducible, that is, there is a submodule that is itself a
representation of the Virasoro algebra.  Such a submodule is
generated from a highest-weight state $|\delta \ket$, such that 
\begin{align}
\label{singular}
  L_n |\delta \ket =0 ~~~~~~~~~~~~n>0, 
\end{align}
although this state is also of the 
form (\ref{state}).  Such a state generates its own Verma module and
is called a {\it singular state}.  In the case of minimal
models \cite{bpz} it is also called a {\it null state}, since its norm defined with respect
to the inner product is
\begin{align}
  \bra \delta | \delta \ket = 0.  
\end{align}

Consider two partitions of level $l$: $\sum \lambda_k k = \sum \lambda'_k k
= l$.  Act on $|\Delta,  \{ \lambda \} \, \ket$ with 
$L_{k_1}^{\lambda'_1} ... L_{k_n}^{\lambda'_n}$. The result is proportional to $|\Delta \ket$:
\begin{align}
   L_{k_1}^{\lambda'_1} ... L_{k_n}^{\lambda'_n} \, |\Delta,  \{ \lambda \} \, \ket
 = M^{(l)}_{\{ \lambda'\}, \{ \lambda\}} \, |\Delta \ket ,
\end{align}
where $ M^{(l)}_{\{\lambda'\}, \{ \lambda\}}$ is a polynomial in
$\Delta$ and $c$.  If $M^{(l)}$ has a zero eigenvalue, then there exists a
singular state (\ref{singular}) at level $l$, and the representation on
the subspace generated by (\ref{state}) is reducible.

There exists a general formula, due to Kac \cite{kac1,kac2}, for the determinant of the
$M^{(l)}$, the {\it Kac determinant}: 
\begin{align}
  \det M^{(l)} = \alpha_l \prod_{\substack{ r, s \geq 1\\ rs \leq l}}
  [\Delta - \Delta_{r,s} (c)]^{p(l-rs)},
\end{align}
where $p(l-rs)$ is the number of partitions of the integer $l-rs$
and $\alpha_l$ is a positive constant independent of $\Delta$ and $c$.

The function $\Delta_{r,s} (c)$ can be written as
\begin{align}
    \Delta_{r,s} = \frac{c-1}{24} + \frac{1}{4} \,  [ \, r \alpha_+ +
    s \alpha_- ]^2, 
\end{align}
with
\begin{align}
  \alpha_{\pm} = \frac{\sqrt{1-c} \pm \sqrt{25-c}}{\sqrt{24}}.
\end{align}
When $\Delta = \Delta_{r,s}$, there exist singular vectors with dimensions 
\begin{align}
\label{rs}
  \Delta_{\delta} = \Delta_{r,s} + rs.
\end{align}

In Liouville field theory the Hartle-Hawking states corresponding to
vertex operators (\ref{construct}) have negative conformal weights 
(\ref{negativeweight}), which can be put in Kac form
\begin{align}
\label{kac}
  \Delta_{J, \hJ} = \frac{c-1}{24} + \frac{1}{4} \,  [2 \hat{J} \alpha_+ + 2 J \alpha_- ]^2, 
\end{align}
where in this case
\begin{align}
  \alpha_+ = i \frac{\sqrt{2}}{\tgamma}, ~~  \alpha_- = i \frac{\tgamma}{\sqrt{2}}.
\end{align}
$J$ and $\hat{J}$ are related to the spins $j$, $\hat{j}$ of the
quantum algebra $ U_{q}(sl_2) \odot U_{\hq}(sl_2)$ by
\begin{align}
  J= j+1/2, ~~ \hat{J} = \hat{j}+ 1/2.
\end{align}
We see that at level $l= 4 J \hat{J}$ there are singular states with $\Delta_{\delta} =
\Delta_{J, \hJ}+l$.

\section{Conformal Ward identity}
\label{ward identity}
\setcounter{equation}{0}

We will see in Section \ref{norm} that just as in minimal models, the singular states 
discussed in Section \ref{null} decouple from the conformal families.  In order to understand this and
to define the norm of these decoupling states, we need to
understand the Ward identity in quantum Liouville theory for two-point
correlation functions.  We will follow the perturbative treatment of
the geometrical approach to two-dimensional quantum gravity by Takhtajan 
\cite{takhtajan1,takhtajan2,takhtajan3,takhtajan4,takhtajan5}, which has a deep connection with the
uniformation problem going back to Poincar\'e's theorem \cite{poincare}. 

The correlation functions of the vertex operators of Liouville field
theory are represented by functional integrals with the Liouville 
action over all Riemannian metrics in a given conformal class with 
prescribed conical singularities at inserted points. 
It was first observed by Polyakov (see \cite{takhtajan5} for details) that 
at the semi-classical level the Ward identities of the quantum
Liouville theory establish a non-trivial relation between the accessory parameters and the 
critical value of the Liouville action functional. 
In the series of work by
Takhtajan \cite{takhtajan1,takhtajan2,takhtajan3,takhtajan4,takhtajan5}, 
for the region $c>1$, the
validity of BPZ conformal Ward identity for $n \ge 3$ puncture operators was
proven perturbatively, with the quantum central charge $c= 1+ 12
/\gamma^2$, in agreement with the result of the algebraic approach
(\ref{quantum_c}). 

The proof can be easily generalized to the
case of $n\ge3$ conical singularities\cite{tz}.  However, we shall see that in the
case of two conical singularities, the Ward identity takes a 
different form, which is related to the geometrical property of a
sphere with two conical singularities.

\subsection{Facts about the sphere with two conical singularities}
\label{sphere}

We will need the following facts about the Riemann sphere with two
conical singularities.

Let $X$ be a Riemann sphere with two conical singularities at $z=0$
and $z=\infty$.  Following Troyanov \cite{troyanov}, a (conformal) metric $ds^2$ on a Riemann surface $S$
has a {\it conical singularity} of order $\beta$ ($\beta$ a real number
$>-1$) at a point $p \in S$ if in some neighbourhood of $p$
\begin{align}
  ds^2 = e^{\varphi} |dz|^2,  
\end{align}
where $z$ is a coordinate of $S$ defined in this neighbourhood and
$\varphi$ is a function such that 
\begin{align}
\label{asym}
  \varphi(z) - 2 \beta \log |z- z(p)|
\end{align} 
is continuous at $p$.

A { \it projective connection} $\ueta(z)$ on a Riemann surface
$S$ is defined as a rule which associates to each local uniformizer $z$ on $S$ a
meromorphic quadratic differential 
\begin{align}
  \ueta(z) = \uT(z) dz^2
\end{align}
defined in the domain of $z$, in such a way that under a
holomorphic change of coordinates
\begin{align}
  \ueta(w) &= \ueta(z) +  \{ z,w \} dw^2 , \\
  \uT (w(z)) &= \uT(z) \Big( \frac{dz}{dw} \Big )^2 + \{ z,w\}, \nonumber
\end{align}
where $\{,\}$ denotes the Schwarzian derivative:
\begin{align}
  \{ f,w \} = \frac{f^{'''}}{f'} - \frac{3}{2} \frac{f^{''}}{f'}.
\end{align}
The projective connection $\ueta$ has a regular singularity of weight
$\uDelta$ if 
\begin{align}
  \ueta = \big ( \frac{\uDelta}{z^2} + \frac{d}{z} + \uT_1 (z) \big ), 
             ~~~ \uT_1 ~ \text{holomorphic},
\end{align}
where $z$ is a uniformizer at $p$ such that $z(p)=0$.  This definition
of weight is independent of the choice of uniformizer.

If $ds^2 = e^{\varphi} |dz^2|$ is a (conformal) metric of
constant curvature on $S$ with conical singularities of order
$\beta_1, \beta_2, ...\beta_n$ at $p_1, p_2,... p_n$, then 
\begin{align}
  \uT(z) = \varphi_{zz} - \frac{1}{2} \varphi_z^2,
\end{align}
where $\varphi$ is a solution of Liouville equation with constant
positive curvature.


Let $\ueta$ be a projective connection on $S^2$ with
regular singularities at $z=0$ and $z=\infty$.  Then we have (in the
standard coordinate $z$):
\begin{align}
\label{proposition}
  \ueta(z) = \frac{\uDelta}{z^2} dz^2, ~~~ \uDelta \in \C. 
\end{align}
In particular, both singularities have the same order $\beta$
and weight 
\begin{align}
\label{weight}
  \uDelta = - \frac{\beta (\beta + 2)}{2}.
\end{align}

\subsection{Correlation functions defined on a sphere with two conical singularities}
\label{correlation}

For the case of two vertex operators, the correlation function is given
by functional integral that diverges, so we need to define the functional integral for
fixed area $A = \int e^{\varphi} d^2 z$, following \cite{onofri,zamolodchikov, seiberg}:
\begin{align}
  \bra  X \ket =  \int dA \, \bra X \ket_A e^{- A / 2 \pi \gamma^2 }, 
\end{align}
where the functional integral for fixed area $A$ is defined as
\begin{align}
\label{area}
  \bra  X \ket_A  = \X{\int} {\mathcal D} \varphi
                  e^{- \frac{1}{2 \pi \gamma^2} I_L^{(A)}} \delta \big
                  ( \int d^2 z e^{\varphi} -A \big ).
\end{align}
Here ${\mathcal C} (X)$ denotes the class of smooth conformal
metrics on $X$ with one conical singularity at $z=0$, another at
$z=\infty$, both have asymptotics (\ref{asym}).  These conditions imply that the
Liouville action with fixed area diverges.  A properly regularized Liouville
action  $I^{(A)}_L$ \cite{takhtajan1,takhtajan2,takhtajan3,takhtajan4,takhtajan5,krasnov} 
with fixed area contains boundary terms around
singularities,
\begin{align}
\label{action}
  I^{(A)}_L = \lim_{\epsilon \rightarrow 0} \Big \{ \int_{X_{\epsilon}} |\partial
                  \varphi|^2  d^2 z + (2 \pi -2) \log \epsilon \Big \},
\end{align}
where $X_{\epsilon}= X \backslash \{ r_1 < \epsilon \} \bigcup  \{ r_2 > 1 /\epsilon \}$.

Similarly, correlation functions of the energy-momentum tensor in the
presence of conical singularities are defined by
\begin{align}
  \bra \,  \prod^{k}_{i=1} T(z_i) \prod^{l}_{j=1} \bar{T}(\bar{w}_j) X
    \,\ket \equiv 
  \int dA \, \bra \,  \prod^{k}_{i=1} T(z_i) \prod^{l}_{j=1} \,
    \bar{T}(\bar{w}_j) X   \, \ket_A e^{- A / (2 \pi \gamma^2) } ,
\end{align}
where
\begin{align}
\label{T_area}
  \bra I \ket_A \, &\equiv \, \bra   \, \prod^{k}_{i=1} T(z_i) \prod^{l}_{j=1} \bar{T}(\bar{w}_j) X
     \, \ket_A  \nonumber \\
  &= \X{\int} {\mathcal D} \varphi 
     \prod^{k}_{i=1} T(z_i) \prod^{l}_{j=1} \bar{T}(\bar{w}_j) 
                  e^{-1/(2 \pi \gamma^2) \int d^2 z |\partial
                  \varphi|^2 } \delta \big ( \int d^2 z e^{\varphi} -A  \big )
\end{align}
As in standard quantum field theory, define the normalized connected correlation function
\begin{align}
\label{normalized}
  \dbra I \dket_A \equiv \dbra \prod^{k}_{i=1} T(z_i) \prod^{l}_{j=1} \bar{T}(\bar{w}_j) X
  \dket_A 
\end{align}
by the following inductive formula:
\begin{align}
  \dbra I \dket_A = \frac{\bra I \ket_A}{\bra X \ket_A}
                  -\sum^{k+l}_{r=2} \sum_{I = I_1 \bigcup ...I_r}
                    \dbra I_1 \dket_A... \dbra I_r \dket_A,
\end{align}
where the summation goes over all representations of the set $I$ as a
disjoint union of the subset $I_1...I_r$.  For example, 
\begin{align}
\label{one}
  \dbra  T(z) X  \dket_A \, &= \,  \frac{\bra  T(z) X  \ket_A}{\bra X
  \ket_A}, \\
\label{two}
  \dbra  T(z) T(w) X  \dket_A \, &=  \, 
  \frac{\bra  T(z) T(w) X \ket_A}{\bra  X \ket_A } + \dbra  T(z) X \dket_A
  \dbra T(w) X \dket_A .
\end{align}

The generating functional for correlation
functions of the stress-energy tensor (\ref{normalized}) is introduced by the following
expression:
\begin{align}
  \cZ (\mu, \bmu; X)_A = \frac{Z (\mu, \bar{\mu}; X)_A}{\bra
  X \ket_A},
\end{align}
where 
\begin{align}
\label{generate_Z}
  Z (\mu, \bar{\mu}; X)_A 
  = \X{\int} &{\mathcal D} \varphi \,\delta \big
                  ( \int d^2 z e^{\varphi} -A \big )  \times \non \\ 
     &\times \exp \Big[- \frac{1}{2 \pi \gamma^2} I^{(A)}_L + \text{p.v.} \int_{X}
       (T(\varphi) \mu + \bT(\varphi) \bmu) \Big ]  ,
\end{align}
where the external sources are represented by Beltrami differentials
$\mu$ on the Riemann surface $X$.  The generating functional for the normalized connected multipoint
correlation function is defined in a standard fashion:
\begin{align}
  \frac{1}{\gamma ^2} {\mathcal W} (\mu, \bar{\mu}; X)_A =
  \log {\mathcal Z} (\mu, \bar{\mu}; X)_A, 
\end{align}
so that 
\begin{align}
\label{generating}
  \gamma^2  \dbra  \prod^{k}_{i=1} T(z_i) \prod^{l}_{j=1} \bar{T}(\bar{w}_j) X
    \dket_A = \frac{\delta^{k+l} {\mathcal W}(\mu, \bar{\mu}; X)_A}
            {\delta \mu(z_1)...\delta \mu(z_k)\delta 
    \bar{\mu}(\bar{w}_1) \delta \bar{\mu}(\bar{w}_l)}  
    \Bigg\vert_{\mu =\bar{\mu} =0}.
\end{align}

\subsection{Semi-classical Ward Identity for two-point functions}
\label{ward}

In the semi-classical limit the correlation function is dominated by the
contribution from the solution $\vpc$ to the classical
Liouville equation.  Before evaluating the functional integral, first
we calculate the classical solution $\vpc$ around which we will do the expansion.
Let us look for classical solution corresponding to a Riemann sphere $S^2$
with two conical singularities at $z=0, \infty$, both of order
$\beta$.  The Gauss-Bonnet formula for such a surface with positive constant
curvature gives 
\begin{align}
  \frac{K}{2 \pi} A  = \chi(S^2) + \sum_{i=1}^2 \beta_i,
\end{align}
where $K$ is the Gaussian curvature, and $\chi = 2$ is the Euler
characteristic of $S^2$.  $\vpc$ is then the solution of the Liouville
equation with constant positive curvature 
\begin{align}
  \partial \bar{\partial} \varphi 
  = -\frac{K}{2}e^{\varphi}
  = -\frac{\pi}{A} \big (2 + \sum_{i=1}^2  \beta_i \big) e^{\varphi}.
\end{align}
We will choose $K=1$ and fixed area $A_0= 2 \pi (2 + \sum
\beta_i)$, but any choice of $A$ can be realized by a shift in
$\varphi$.  With this choice the Liouville equation for the classical solution becomes
\begin{align}
  \partial \bar{\partial} \varphi = -\frac{1}{2} e^{\varphi}.
\end{align}
In order to expand the Liouville action around the classical solution,
first let 
\begin{align}
  \varphi = \varphi_{\text cl} + \delta \varphi.
\end{align}
The expansion of the Liouville action (\ref{action}) is then
\begin{align}
  I^{(A_0)}_L(\varphi_{\text cl} + \delta \varphi) 
  = I^{(A_0)}_L(\vpc) 
    - \underset{X}{\int} (\delta \varphi) (L_0 + 1/2) (\delta \varphi) \, d \rho 
    - \sum^{\infty}_{k=3} \frac{1}{k!} \underset{X}{\int}\, (\delta \varphi)^k \, d \rho,
\end{align}
where $d\rho = e^{\vpc} d^2 z$ is the volume form of the metric, and
\begin{align}
\label{laplacian}
  L_0 = e^{-\vpc} \partial \bpartial 
\end{align}
is the Laplacian operator on $X$.  The inverse of $(2 L_0 +1)$ is given
by the Green's function $G(z,z')$.  The
functional integral (\ref{area}) now becomes 
\begin{align}
  \bra  X \ket_{A_0}  
  = \X{\int} &{\mathcal D} [\delta \varphi] \,\delta \big ( \int d^2 z \,
    e^{\varphi} -A_0 \big ) \times \non \\
      & \times e^{ - \frac{1}{2 \pi \gamma^2} [I_L^{(A_0)}(\varphi_{\text{cl}}) 
    - \underset{X}{\int} (\delta \varphi) (L_0 + 1/2) (\delta \varphi) d \rho 
    - \sum^{\infty}_{k=3} \frac{1}{k!} \underset{X}{\int} (\delta \varphi)^k d
                  \rho ]} .
\end{align}
We also need to expand $T(\varphi)$ and $\bT(\varphi)$ around the
classical solution.  The  expansion of the generating functional
(\ref{generate_Z}) then reads
\begin{align}
   Z (\mu, \bar{\mu}; X)_{A_0} 
    = & ~\exp \big [ -\frac{1}{ 2 \pi \gamma^2}
       I^{(A_0)}_L(\vpc) + \text{p.v.} \int_X (\Tc \, \mu +
        \bTc \, \bar{\mu} ) d^2 z \big ] \times \nonumber \\
     &\times \exp \Big [ \frac{1}{2 \pi \gamma^2}
      \sum^{\infty}_{k=3} \frac{1}{k!} \int_X (\gamma^2
      \frac{\delta}{\delta \xi})^k d \rho \Big ] \times \\
     &\times \Bigg ( \X{\int} {\mathcal D}
     [\delta \varphi] \,\delta \big [ \int d^2 z
     e^{\vpc}(e^{\delta \varphi} -1) \big ] \times \non \\
      & \times \exp \Big \{\frac{1}{2 \pi \gamma^2} \underset{X}{\int} \big 
      [(\delta \varphi)' \, (L_0 + 1/2 - \pi e^{-\vpc}(\partial \mu
     \partial + \bar{\partial}\, \bar{\mu} \,\bar{\partial})) \, (\delta \varphi)' ] \,
      d \rho \Big \} \times  \nonumber \\
      & ~~~~~~~~~~\times \exp \Big \{- \frac{ \pi}{\gamma^2} \underset{X}{\int} \xi \, G
      \big [ 1 - 2 \pi G e^{-\vpc}(\partial \mu
     \partial + \bar{\partial}\, \bar{\mu}\, \bar{\partial})  
     \big ]^{-1} \xi \, d \rho \Big \} \Bigg )   \nonumber 
\end{align}
where 
\begin{align}
  \xi = e^{-\vpc}(\omega + \bar{\omega}),  ~~~~~~
  \omega = \mu_{zz} + (\vpc)_z \mu_z + (\vpc)_{zz} \mu,
\end{align}
and
\begin{align}
  \delta \varphi' = \delta \varphi + \frac{ \pi \xi}{L_0 + 1/2 - \pi e^{-\vpc} (\partial \mu
     \partial + \bar{\partial}\, \bar{\mu} \,\bar{\partial})}.
\end{align}

Now look at the tree level value of $\dbra T(z) X \dket_{A_0}$ and $ \dbra T(z) T(w)X
\dket_{A_0}$, which are defined in (\ref{one}) and (\ref{two}).
According to the definition (\ref{generating}),
\begin{align}
  \dbra T(z) X \dket_{A_0} = \frac{1}{\gamma^2} 
   \frac{\delta {\mathcal W}(\mu,\bmu;X)_{A_0}}{\delta \mu (z)} \Big \vert_{\mu=0}. 
\end{align}
It is clear that 
\begin{align}
\label{one-tree}
  \dbra T(z) X \dket_{A_0 - \text{tree}} = T_{A_0 - \text{cl}} (z) 
  = \frac{1}{\gamma^2} \frac{\uDelta}{z^2},
\end{align}
where $\uDelta$ is defined as in (\ref{weight}), due to the form of
the projective connection  (\ref{proposition}) on $X$.  

This is of a different form from the case of $n \ge 3$ conical singularities
\cite{takhtajan1,takhtajan2,takhtajan3,takhtajan4,takhtajan5}.  For a Riemann sphere with $n \ge 3$ conical
singularities of deficit angle $\theta = 2 \pi (1- \alpha_i), i = 1,
\cdots,n$, with
\begin{align}
  \alpha_i <1, ~~~~~ \sum^{n-1}_{i=1} \, \alpha_i >2, 
\end{align}
there exists \cite{picard, lichtenstein, troyanov2} a unique conformal metric $ds^2 =
e^{\varphi} |dz|^2$ of constant curvature $-1$, where $\varphi$ is a
smooth function satisfying the Liouville equation (\ref{eom3}) with
certain asymptotics near the singular points.  At the tree level \cite{tz}
\begin{align}
\label{one-tree_2}
  \dbra T(z) X \dket_{\text{tree}} = T_{\text{cl}} (z)  
  = \sum_{i=1}^{n-1} \Big ( \frac{h_i}{(z- z_i)^2} + \frac{c_i}{z-z_i} \Big ),
\end{align}  
where $h_i = \alpha_i (2 - \alpha_i)$.  The complex number $c_i$ are
the famous {\it accessory parameters}, which are uniquely determined
by the singular points $z_1, \cdots, z_n$ and the set of orders
$\alpha_i$.

For the connected Ward identity 
\begin{align}
  \dbra T(z) T(w)X \dket_{A_0} = \frac{1}{\gamma^2}
   \frac{\delta^2 {\mathcal W}(\mu, \bmu;X)_{A_0}}{\delta \mu (z) \delta \mu(w)} \Big \vert_{\mu=0}, 
\end{align}
at the tree level, the only terms of order $1/\gamma^2$ come from  
\begin{align}
 \exp \Big [ \frac{1}{2 \pi \gamma^2}
         \sum^{\infty}_{k=3} \frac{1}{k!} \int_X (\gamma^2
         \frac{\delta}{\delta \xi})^k d \rho \Big ]
\end{align}
acting on the first term in the expansion of the integral
\begin{align}
\exp \Big \{- \frac{ \pi}{\gamma^2} \underset{X}{\int} \xi \, G
      \big [ 1 - 2\pi G e^{-\varphi_{\text{cl}}}(\partial \mu
     \partial + \bar{\partial}\, \bar{\mu}\, \bar{\partial})  
     \big ]^{-1} \xi \, d \rho \Big \} \Bigg ).   
\end{align}
We obtain the following expression
\begin{align}
\label{two-point}
 \dbra T(z) T(w)X \dket_{A_0-\text tree} = -\frac{2 \pi}{\gamma^2} 
        {\mathcal D}_z {\mathcal D}_w G(z,w).
\end{align}
where ${\mathcal D}_z = \partial_{z}  \partial_{z}- (\vpc)_z \partial_z$.

\subsection{Ward identity on the sphere with two conical singularities}
\label{wi on sphere}

We will derive the Ward identity for the insertion of two vertex operators 
$e^{-\varphi / \gamma \tgamma}$ at $z=0, \infty$, which classically corresponds to $\beta
=1$.  In this case two spheres are cut along a geodesic joining the
north pole to south pole and then glued together \cite{troyanov}, so that
the tangent cones at the insertion point $z=0, \infty$ are of
angles $\theta = 4 \pi$.  There exists a map from  $z$-plane $X$
to  $\zeta$-plane $S^2$,
\begin{align}
\label{map}
  \zeta = z^2,
\end{align}
which is a covering $f: S^2 \rightarrow S^2$.  

The kernel of the integral operator $(2 L_0+1)$ is given by the
Green's function $G(z,z')$, which satisfies on $X$ the PDE
\begin{align}
  2 G_{z \bar{z}}(z,z') + \exp[\varphi_{\text{cl}}] G(z,z') = \delta(z-z').
\end{align}

We first study the Green's function $\mathcal G(\zeta, \zeta')$ on the Riemann sphere $S^2$ of
radius $R=1$,
on which the Laplacian (\ref{laplacian}) is of the form
\begin{align}
  L_0 = \frac{1}{4} (1+ \zeta \bar{\zeta})^2 \, \partial_{\zeta} \partial_{\bar{\zeta}}.
\end{align}
Define the point-pair invariant
\begin{align}
\label{distance}
  u= - \frac{(\zeta - \zeta')(\bar{\zeta} - \bar{\zeta}')}
    {( 1+ \zeta \bar{\zeta} )( 1+ \zeta' \bar{\zeta'} )}.
\end{align}
Then the explicit solution for the
Green's function is
\begin{align}
\label{gf}
  \mathcal G = - \frac{1}{2 \pi} (2u +1) \ln(\frac{u+1}{-u}) + \frac{1}{\pi},
\end{align}
with the desired asymptotic behavior.  Note that here $u \in (-1,0)$.

Using the map (\ref{map}) to obtain $G(z,z')$ on $X$ from $\mathcal G(\zeta, \zeta')$, then using
(\ref{two-point}), a straightforward calculation reveals that for $\beta =1$,  at the tree level, 
\begin{align}
   \dbra T(z) T(w)X \dket_{A_0} = \frac{6}{\gamma^2} 
     \Big [ \frac{1}{(z-w)^4} - \frac{1}{2(z-w)^2 w^2}  
     + \frac{1}{2 (z-w) w^3} \Big ],
\end{align}
which combined with (\ref{one}), (\ref{two}) and (\ref{one-tree}), gives rise to the expression 
\begin{align}
\label{two-tree}
  \bra T(z) T(w)&X \ket_{A_0} = \, \frac{c_{\text{cl}}/2}{(z-w)^4} \bra X
  \ket_{A_0} +\non \\
  &+ \Bigg \{ \frac{2}{(z-w)^2} + \frac{1}{z-w}
  \frac{\partial}{\partial w} + \sum^{2}_{i=1} 
  \frac{\Delta}{(z-z_i)^2} \Bigg \}  \, \bra T(w) X \ket_{A_0},
\end{align}
where $c_{\text{cl}} = 12/ \gamma^2$ is the classical central charge,
and $\Delta = -3 /2 \gamma^2  =\uDelta/ \gamma^2$ is the conformal
weight of the vertex operator as $\tgamma \rightarrow 0$.

In the standard conformal Ward identity obtained from the operator
product expansion (OPE) \cite{bpz},
\begin{align}
  \bra T(z) T(w)X \ket &= \frac{c/2}{(z-w)^4} \bra X \ket 
  + \, \Bigg \{ \, \frac{2}{(z-w)^2}+ \non \\
  &+ \frac{1}{z-w}
  \frac{\partial}{\partial w} + \sum^{n}_{i=1}  \Big (
  \frac{\Delta}{(z-z_i)^2}  + \frac{1}{z-z_i} \frac{\partial}{\partial
   z_i} \Big ) \Bigg \}  \, \bra T(w) X \ket,  
\end{align}
where the terms after the summation sign come from the OPE of the vertex
operators, while the terms before that are due to the OPE of the
energy-momentum tensor itself.  Comparing with (\ref{two-tree}),
we see that the contributions from the OPE of the $T$'s remain the
same; all difference comes in from the fact that 
\begin{align}
  \bra T(z) X \ket_{A_0} =  \dbra T(z) X \dket_{A_0} \bra X \ket_{A_0} 
   =  \sum^{2}_{i=1} \frac{\Delta}{(z-z_i)^2} \, \bra X \ket_{A_0}
\end{align}
is different from the standard form
\begin{align}
   \bra T(z) X \ket = \sum^{n}_{i=1}  \Big (
  \frac{\Delta}{(z-z_i)^2}  + \frac{1}{z-z_i} \frac{\partial}{\partial
   z_i} \Big ) \, \bra X \ket.
\end{align}
This difference, in turn, is the consequence of the difference between (\ref{one-tree})
and (\ref{one-tree_2}).

In Section \ref{vacuum}, we saw that the $SL(2,C)$-invariant
vacuum does not exist in Liouville field theory.  Thus it is not
surprising to have a Ward identity that is different from the
standard form, since we do not have the usual operator-state
correspondence; instead we treat each 
Hartle-Hawking state as a ground state, and use the OPE of energy-momentum
tensor to define $\bra  \prod^{k}_{i=1} T(z_i) X \ket_{A_0}$ from $\bra
T(z) X \ket_{A_0}$.  

In the BPZ formulation \cite{bpz}, all correlation functions of secondary fields are given by
differential operators acting on those of primary states.  Here we
define differential operators $\cL_{-n}$, consistent with our semi-classical result
\begin{align}
\label{diffop}
  \bra (L_{-n} V)(w) V (z) \ket_{A}  = \cL_{-n} \bra V(w) V(z) \ket_{A}, ~~~~~
  (n \ge 1),
\end{align}
with a modified definition 
\begin{align}
  \cL_{-n} = \frac{(n-1) \Delta}{(w-z)^2}.
\end{align}
In particular,
\begin{align}
\label{1_2}
  \cL_{-1} = 0, ~~~~~\cL_{-2} =\frac{\Delta}{(w-z)^2}. 
\end{align}

\section{Decoupling states with nonzero norm}
\label{norm}
\setcounter{equation}{0}

\subsection{The norm of the singular states}
\label{norm_singular}

We now proceed to define the norm of the singular states described in
Section \ref{null}.  Because of the nonstandard form of the Ward
identity of two-point functions and the differential operators
(\ref{diffop}), we will show by an explicit example
here that the norm of such a singular state  is in fact not zero,
unlike the familiar case of the minimal models. 

A simple example of such a singular state is at level $l=2$ of the
Hartle-Hawking state $|\Delta \ket$ constructed from the operator $V=e^{ -\tgamma \phi/2 }$, 
\begin{align}
  | \delta \ket = [L_{-2} + \eta L_{-1}^2] \, |\Delta>, 
\end{align}
where 
\begin{align}
  \eta = - \frac{3}{2 (2 \Delta +1 )}.
\end{align}
The norm of such a state can be evaluated by gluing together two disks
with operator $\delta(z, \bz)$ at the center of each disk.

The norm of the singular state $\bra \delta |  \delta \ket$
is then related to the correlation function by
\begin{align}
   \bra \delta|  \delta \ket 
  &= \lim_{w, \bar{w} \rightarrow \infty} 
     \bar{w}^{2 (\bDelta + 2)} w^{2 (\Delta + 2)}
     \bra \delta (\bar{w},w)  \delta (0,0) \ket \non \\
  &= \lim_{w, \bar{w} \rightarrow \infty} 
     \bar{w}^{2 (\bDelta + 2)} w^{2 (\Delta + 2)}
     \bra (L_{-1}^2 + \eta L_{-2}) V(\bar{w},w) 
      (L_{-1}^2 + \eta L_{-2}) V(0,0) \ket, \nonumber\\
  &=  \lim_{w, \bw \rightarrow \infty} \eta^2 \Delta^2  \bw^{2 \bDelta}
      w^{2 \Delta }  \int dA \, e^{- A / (2 \pi \gamma^2) } 
    \bra V(\bar{w},w)  V(0,0) \ket_A  \nonumber\\
  &=  \lim_{w, \bw \rightarrow \infty} \eta^2 \Delta^2  \bw^{2 \bDelta}
      w^{2 \Delta } \bra V(\bar{w},w)  V(0,0) \ket.  
\end{align}
The two-point function are expressed in terms of gamma functions as \cite{do,fzz}
\begin{align}
   & \bra e^{\alpha \phi(w)} e^{\alpha \phi(0) } \ket \non \\
  = &\Big [ 4 \pi \frac{ \Gamma(\tgamma^2/2)}{\Gamma(1- \tgamma^2/2)} 
    \Big ]^{2(\frac{1}{\gamma}- \alpha)/ \tgamma} \frac{2}{\tgamma^2} \,
    \frac{\Gamma( \tgamma \alpha - \tgamma^2 /2) \,
          \Gamma (2 \alpha/\tgamma - 2/ \tgamma^2 -1)}
         {\Gamma( 1- \tgamma \alpha + \tgamma^2 /2) \,  
          \Gamma (2 - 2 \alpha/\tgamma + 2/ \tgamma^2 )}
     \frac{1}{(w \bar{w})^{2 \Delta}}.
\end{align}

\begin{figure}
  \psfrag{0}{\tiny $0$}
  \psfrag{inf}{\tiny $\infty$}
  \centering{\includegraphics [scale=1]{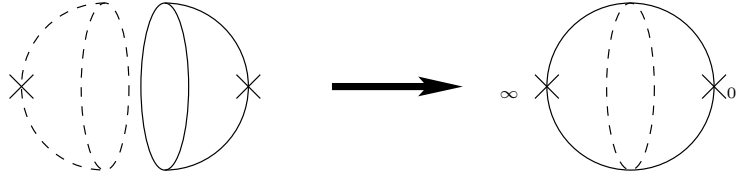}}
  \caption{Norm of Hartle-Hawking state}
\end{figure}

\noindent Using the two-point function for $V=e^{- \tgamma \phi/2 }$, we
  obtain
\begin{align}
\label{nonzero_norm}
  \bra \delta|  \delta \ket 
  = & \eta^2 \Delta^2 \Big [ 4 \pi \frac{ \Gamma(\tgamma^2/2)}{\Gamma(1- \tgamma^2/2)} 
    \Big ]^{2(\frac{1}{\gamma}- \alpha)/ \tgamma} \frac{2}{\tgamma^2} \,
    \frac{\Gamma( -\tgamma^2) \, \Gamma (-2/ \tgamma^2 -2)}
         {\Gamma( 1+ \tgamma^2) \, \Gamma (3 + 2/ \tgamma^2 )}.
\end{align}

\subsection{Decoupling of the singular states}
\label{decoupling}

In minimal models, the representation of the Virasoro algebra $V_{\Delta}$ is
irreducible unless the dimension $\Delta$ takes values in Kac table.
The singular vector $| \delta \ket$ is orthogonal to any state of
$V_{\Delta}$ and has zero norm. Thus such a singular state is also
called a null state, and all its descendants are also null states
since their norms are proportional to  $\bra \delta| \delta \ket$.

That such a null state in minimal models is orthogonal to the whole Verma module
translates, in the field language, into the vanishing of the
correlator $\bra \delta X \ket$, where $X$ is a string of local
fields:$X \equiv  \phi_(z_1) \cdots \phi_N(z_N)$.  This implies
certain differential equation for $\bra \delta X
\ket$.  In the example for level-$2$ null state the differential
equation is
\begin{align}
\label{decoupling_1}
  \{ \cL_{-2} + \eta \cL^2_{-1}  \} \bra \phi(z)X \ket  =0.
\end{align}
Or more explicitly
\begin{align}
\label{decoupling_2}
   \Bigg\{\sum_{i=1}^N \Bigg [ \frac{1}{z-z_i} \frac{\partial}{\partial
   z_i} + \frac{\Delta_i}{(z-z_i)^2}\Bigg ] +\eta\, 
   \frac{\partial^2}{\partial z^2}  \Bigg \} \bra \phi(z)X \ket  =0,
\end{align}
where $\Delta_i$ is the dimension of the primary field $\phi_i$.

Now look at the singular states in the Liouville field theory.  We see
that eqn. (\ref{decoupling_2}) remains valid for $N \ge 2$, since in
this case the Ward identity is of the standard form.  For
$N=1$ eqn. (\ref{decoupling_1}) took a different form  because of
$(\ref{1_2})$.  Yet because of the $\delta$-function in the two point
function
\begin{align}
  \bra e^{\alpha \phi(w)} e^{\beta \phi(0)} \ket  
  \sim  \frac{\delta(\alpha - \beta)}{(w \bw)^{2 \Delta}},
\end{align}
the singular state is still orthogonal to the whole Verma modules but
itself - that is, it is a decoupling state with nonzero norm (\ref{nonzero_norm}).

\section{BTZ Black hole entropy}
\label{counting}
\setcounter{equation}{0}

Now let us look at the states in Liouville field theory that are candidates for the
BTZ black hole state counting. 

As we have seen in Section \ref{sec:group} and \ref{construction},  the
deformation parameters of the quantum algebra 
$U_q(sl_2) \odot U_{\hq}(sl_2)$ in Liouville field theory are
\begin{align}
  \begin{cases}
  ~q=\exp(i \pi \tgamma^2/2), ~&~~~ p = 4/\tgamma^2,
  \non \\
  ~\hq = \exp(i 2 \pi/ \tgamma^2), ~&~~~ \hp = \tgamma^2.
  \end{cases}
\end{align}

Let us consider the case in which $4/ \tgamma^2 =2N+1$,
an odd integer, so
\begin{align}
   q=e^{i 2 \pi /(2 N+1 )}, ~~~~~~\hq =e^{i (N+ 1/2) \pi} .
\end{align}
Accordingly the Hartle-Hawking states with negative conformal weight
$\Delta_{J, \hJ}$ given by (\ref{kac}) have spin 
\begin{align}
  J &= 1/2, 1, ..., 2/ \tgamma^2,\nonumber \\
  \hat{J} &= 1/2, 1. 
\end{align}
The conformal families built on these states are reducible, as we have
discussed in Section \ref{null}.  The
decoupling singular states in these reducible Verma modules $\Delta_{J,
  \hJ}$ have conformal weights
\begin{align}
  \Delta_{\delta} =  \Delta_{J, \hJ} + 4 J \hJ 
         = \frac{c-1}{24} + \frac{1}{4} \,  [2 \hat{J} \alpha_+ - 2 J \alpha_- ]^2,
\end{align}
and are highest weight states themselves. Yet unlike in minimal models,
the Verma modules built on these decoupling states are irreducible
without further decoupling, since the $rs$ term in eqn. (\ref{rs}) is now negative.
 
In particular, for states with $\omega = \pm1$ and spins
\begin{align}
\label{states}
  \hJ &=\frac{1}{2}, ~~~~~~~~~~
  J= \frac{1}{2}, 1, \cdots, \frac{2}{\tgamma^2},~~~~~~~~~~{\text {and}} \non \\
   ~~~~~~~~~~~~~  \hJ &=1, ~~~~~~~~~~~
  J = \frac{1}{\tgamma^2} - \frac{1}{4}, \cdots,\frac{2}{\tgamma^2},
\end{align}
$\Delta_{\delta}$ takes value between $0$ and $2/ \gamma^2$.
Altogether there are $12/ \tgamma^2$ such states. For 
other spin values $\Delta_{\delta}$ is negative.


As we showed in Section \ref{norm}, unlike the null states of minimal models, these
decoupling states have nonzero norms.  Furthermore, they can have
positive definite norm for certain values of $N$.  Consider for example,
the decoupling state at level $l=2$ of the Hartle-Hawking state
constructed from the operator $V= e^{- \tgamma \phi /2}$.  Its norm
was given in eqn. (\ref{nonzero_norm}), and the sign of the norm is
determined by 
\begin{align}
  \Gamma(-\tgamma^2)\,  \Gamma (-2/\tgamma^2 -2 ) 
  = (-1)^{N+3} \, \frac{2^{N+3} \sqrt{\pi}}{(2N +5)!!}\,  \Gamma(-\tgamma^2).
\end{align}
We can see from the properties of gamma function that for large odd $N$, the
norm (\ref{nonzero_norm}) is positive definite and finite.  

Furthermore, the Verma modules built on these decoupling states are unitary
representations of the Virasoro algebra, since these are
representations with $c > 1, \Delta_{\delta} > 0$.  The proof follows
directly from the Kac determinant; see for example \cite{francesco}.

\begin{figure}
  \psfrag{c>1}{$c>1$}
  \psfrag{Delta>0}{$\Delta>0$}
  \psfrag{Delta=0}{$\Delta=0$}
  \psfrag{Delta<0}{$\Delta<0$}
  \psfrag{|Delta>}{$| \Delta \ket$}
  \psfrag{|delta>}{$| \delta \ket$}
  \centerline{\includegraphics[scale=1]{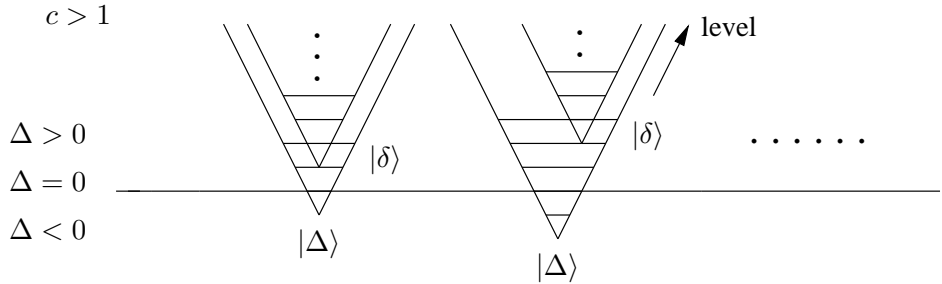}}
  \caption{Verma modules built on $| \Delta \ket$ with decoupling states $| \delta \ket$} \label{fig:structure}
\end{figure}

Thus we propose that the Hilbert space $\cH$ contributing to black hole
entropy is a tensor product of $12/\tgamma^2$ unitary irreducible Verma modules $H$ built on
the decoupling states described in (\ref{states})
with conformal weights $0 < \Delta_{\delta} < 1 / (2 \gamma^2)$:
\begin{align}
  \cH &= \underbrace{H \otimes H \otimes ... \otimes H}. \\
   &\hspace{1.5cm} 12/\tgamma^2 \non
\end{align}
The structure is illustrated in Fig. \ref{fig:structure}.

The conformal weight $\Delta$ is the sum over these $12/\tgamma^2$
sectors.  Because these sectors decouple from each other,  
the asymptotic density of states contributed from each Verma module $H$ 
is equivalent to that of a theory with $12/\tgamma^2$ scalar fields,
\begin{align}
   \ln \rho \, (\Delta) =  2 \pi \, \sqrt{\frac{12}{\tgamma^2} \, \frac{ \Delta}{6}} .
\end{align}
For the semi-classical limit $\tgamma \rightarrow 0$, this coincides
with (\ref{cardy}), the result from Cardy's formula.

\section{Conclusion and discussion}
\label{conclusion}
\setcounter{equation}{0}

In this paper we have followed the canonical quantization approach of Liouville
field theory due to Gervais and his collaborators, which shows an
explicit quantum algebra structure $U_q(sl_2) \odot U_{\hq}(sl_2)$ 
of the quantum Liouville theory.  We considered the vertex operators
that correspond to the graded tensor products of the irreducible representations
of the quantum algebra with positive half-integer spins $j$, $\hj$.  
The corresponding Hartle-Hawking states have negative conformal
weights of Kac form, and the conformal families built on these
highest-weight states are reducible. Yet unlike in minimal models,
the decoupling states are not null states, due to the nonstandard
form of the Ward identity for two-point functions.  We showed that when the deformation
parameter is a root of unity, more specifically, when
$4/\tgamma^2$ is an odd integer, there are natural cut-offs for spins $j$ and
$\hj$.  The conformal families built on the decoupling states with
positive conformal weights, which take value between $0$ and
$2/\gamma^2$,  give rise to the correct Bekenstein-Hawking entropy for BTZ black hole. 

This is a first step towards a thorough understanding of
the microscopic states of the BTZ black hole.  Many questions still
remain to be answered, however.

\begin{itemize}

\item  The derivation of the Ward identity for two-point functions is
  given here only for the tree-level calculation for ``heavy'' vertex
  operators.  It would be interesting to have a nonperturbative
  derivation instead. 

\item We need to understand whether the result holds for general
  values of $\tgamma$.  As a first step, what happens if $4/\tgamma^2$
  is an even integer or a general rational number?  For certain gauge
  groups, there are indications that although Chern-Simon theory is
  well-defined for any coupling constant, but the physical Hilbert
  space becomes finite for rational coupling constants and is
  different from the general case
  \cite{polychronakos1,polychronakos2,hayashi}.

\item  When the deformation parameter is a root of unity, the fusion
  rules of the irreducible representations of the quantum algebra
  $U_q(sl_2)$ necessarily include certain indecomposable
  representations, which may be related to the logarithmic
  operators in the theory.  It will be important to understand the role
  played by these indecomposable representations.  

\item We must still understand the geometric meaning of these decoupling states.

\item It will be interesting to understand the state counting when
  matter fields are coupled to the gravitational field. 

\item Liouville theory may also be obtained near the horizon of an
  arbitrary black hole by dimensionally reducing to the $r-t$ plane.
  Thus the calculation may be extended towards the understanding of
  the gravitational degrees of freedom contributing to an arbitrary
  black hole horizon.

\end{itemize}

\noindent{\large\bf Acknowledgments}\\ 

This work was supported in part by the U.S.\ Department of Energy
under grant number FG02-90ER-40577 and UC Davis Dissertation
Fellowship.  I would like to thank Prof. Steve Carlip for suggesting
this project to me and for his insightful
advice.  I would also like to thank Dr. Krasnov, Prof. Leon Takhtajan and
Prof. Stephen Sawin for kindly answering many of my questions.

\appendix

\section{Appendix}
\label{definition}

The following discussion basically follows \cite{klimyk}.\\

\noindent Let $\K$ stand for a commutative ring with unit. \\

\noindent {\bf Definition 1.} An (associative) {\it algebra} (with unity) is a vector space 
  $\A$ over $\K$ together with two linear maps, $m: \A \otimes \A\rightarrow \A
  $, called the {\it multiplication} or the {\it product}, and 
  $\eta: \K \rightarrow \A$, called the {\it unit}, such that 
  \begin{align}
    m \,  \circ \, (m \otimes \id) &=  m \,  \circ \, (\id \otimes m),  \non \\
    m \,  \circ \, (\eta \otimes \id) = &\, \id  =  m \,  \circ \,  (\id \otimes \eta). 
  \end{align}

\noindent We now dualize this definition by reversing all arrows and replacing all mappings by the corresponding dual ones. \\

\noindent {\bf Definition 2.} A {\it coalgebra} is a vector space $\A$
over $\K$ equipped with two linear mappings, $\Delta: \A \rightarrow \A
  \otimes \A$, called the {\it comultiplication} or the {\it
  coproduct} of $\A$, and $\varepsilon : \A \rightarrow \K$, called
  the {\it counit}, such that 
  \begin{align}
    (\Delta \otimes \id) \circ \, \Delta &= (\id \otimes \Delta) \circ \, \Delta, \non \\
    (\varepsilon \otimes \id) \circ \, \Delta = & \, \id
     =  (\id \otimes \epsilon) \circ \, \Delta. 
  \end{align}

\noindent {\bf Definition 3.} A {\it bialgebra} is a vector space which is an algebra and 
  a coalgebra, that is, the following two conditions are equivalent:
  \renewcommand{\theenumi}{(\roman{enumi})} 
  \begin{enumerate}
   \item $\Delta: \A \rightarrow \A \otimes \A$ and $\varepsilon : \A \rightarrow \K$ 
   are algebra homomorphisms.
   \item $m: \A \otimes \A\rightarrow \A$ and $\eta: \K \rightarrow \A$ are coalgebra 
   homomorphisms.
  \end{enumerate}

\noindent {\bf Definition 4.} A bialgebra $\A$ is called a {\it Hopf algebra} if
  there exists a linear mapping $\CS: \A \rightarrow \A$, called the
  {\it antipode} or the {\it coinverse} of $\A$, such that 
  \begin{align}
     m \, \circ \, (\CS \otimes \id) \circ \, \Delta = \eta \, \circ \, \varepsilon 
     =   m \, \circ \, (\id \otimes \CS) \circ \, \Delta. 
  \end{align}

\noindent The elements of the ring $\C[[h]]$ are formal power series
     $f= \sum^{\infty}_{n=0} a_n h^n$ in an indeterminate $h$ with
     complex coefficient.  Let $V$ and $W$ be vector spaces over
     $\C[[h]]$.  The topology on $V$ for which the set $\{ h^n V +v |
     n \in \N_0 \}$ are a neighborhood base of $v \in V$ is called the
     {\it $h$-adic topology}.  Denote by $V \hotimes W$ the completion
     of the tensor product space $V \otimes_{\C[[h]]} W$ in the $h$-adic topology.\\

\noindent {\bf Definition 5.} An {\it $h$-adic Hopf algebra} $\A$
     is a vector space over $\C[[h]]$ which is complete in the
     $h$-adic toplogy and endowed with $\C[[h]]$-linear mappings 
     $m: \A \hotimes \A\rightarrow \A, \eta: \C[[h]] \rightarrow \A, \Delta: \A \rightarrow \A
     \hotimes \A, \varepsilon : \A \rightarrow \C[[h]]$ and $\CS: \A
     \rightarrow \A$ which satisfy the Hopf algebra axioms with
     $\otimes$ replaced by $\hotimes$.\\

\noindent Let $\tau$ denote the flip operator given by $\tau(a \otimes b) = b
\otimes a$.  Define the coopposite coproduct $\Delta^{\cop} \equiv \tau \circ \Delta$. 

\noindent {\bf Definition 6.} A bialgebra (resp. Hopf algebra) $\A$ is called {\it quasitriangular} if
  there exists an invertible element $R$ of $\A \otimes \A$, such that 
  \begin{align}
     \Delta^{\cop}(a) = R &\Delta(a) R^{-1}, ~~~~~~~~ a \in A \\
     (\Delta \otimes \id)(R) = R_{13} R_{23}, &~~~~~ 
     (\id \otimes \Delta )(R) = R_{13} R_{12},
  \end{align}
where $R_{12}= \sum_i x_i \otimes y_i \otimes 1$, $R_{13}=  \sum_i x_i 
\otimes 1 \otimes y_i$, $R_{23}= \sum_i 1 \otimes x_i \otimes y_i $,
for $R= \sum_i x_i \otimes y_i$.  An invertible element $R \in
\A \otimes \A$ is called a {\it universal R-matrix} of $A$.  A
quasitriangular biagebra (resp. Hopf algebra) with universal
R-matrix $R$ is said to be {\it triangular} if $R_{21}= R_{-1}$, where
$R_{21} = \tau(R) \equiv \sum_i y_i \otimes x_i$.\\

\noindent {\bf Proposition} Let $A$ be a quasitriangular bialgebra
with universal $R$-matrix $R$, then we have 
\begin{align}
  \label{yang} 
  R_{12} R_{13} R_{23} &= R_{23} R_{13} R_{12},\\
  (\varepsilon \otimes \rm id)(R) = 
  &(\varepsilon \otimes \rm id)(R) = 1.
\end{align}
If $A$ is a Hopf algebra, then we also have
\begin{align}
  (S \otimes \id)(R) = R^{-1}, ~~~ (\id \otimes S)(R^{-1}) = R, ~~~ 
  (S \otimes S)(R) =R.
\end{align}
The relation (\ref{yang}) is called the {\it Quantum Yang-Baxter equation}.


\end{document}